\documentclass[12pt]{article}

\usepackage[left=0.8in,right=0.8in,top=1in,bottom=1in]{geometry}
\usepackage{amsmath,amsthm,bbm,amssymb, amsfonts}
\usepackage{mathtools}

\usepackage{graphicx,psfrag,epsf}
\usepackage{enumitem}
\usepackage[round]{natbib}
\usepackage{url} 
\usepackage{color}

\usepackage{booktabs}
\usepackage{siunitx}
\usepackage{array}
\usepackage{rotating}
\usepackage{xr-hyper}
\usepackage{xr}
\usepackage{multirow}
\usepackage[colorlinks, linkcolor=blue, anchorcolor=blue, citecolor=blue]{hyperref}

\usepackage{adjustbox}

\usepackage{url}

\usepackage{multicol}
\usepackage{xcolor}

\DeclarePairedDelimiter\xnorm{\lVert}{\rVert}
\NewDocumentCommand{\norm}{som}
 {\IfBooleanTF{#1}
   {\xnorm*{#3}}
   {\IfNoValueTF{#2}
     {\mathopen{|\mkern-.8mu|}#3\mathclose{|\mkern-.8mu|}}
     {\xnorm[#2]{#3}}%
   }
 }

\makeatletter
\def\breve{\mathpalette\wide@breve}
\def\wide@breve#1#2{\sbox\z@{$#1#2$}%
	\mathop{\vbox{\m@th\ialign{##\crcr
				\kern0.08em\brevefill#1{0.8\wd\z@}\crcr\noalign{\nointerlineskip}%
				$\hss#1#2\hss$\crcr}}}\limits}
\def\brevefill#1#2{$\m@th\sbox\tw@{$#1($}%
	\hss\resizebox{#2}{\wd\tw@}{\rotatebox[origin=c]{90}{\upshape(}}\hss$}
\makeatletter

\DeclareMathOperator*{\argmin}{argmin} 

\newtheorem{theorem}{Theorem}[section]

\newtheorem{prop}{Proposition}
\newtheorem{definition}{Definition}[section]

\def\indep{{\perp \!\!\! \perp}}

\def\gobblestop#1#2{#1}
\def\killstop{%
	\aftergroup\gobblestop
}

\def\thick#1{\hbox{\rlap{$#1$}\kern0.25pt\rlap{$#1$}\kern0.25pt$#1$}}

\def\smbalpha{\boldsymbol{{\scriptstyle{\alpha}}}}

\def\smbalpha{\widehat{\smbalpha}}

\def\hbar{\bar{ h}}

\def\mybox#1{\vskip1mm \begin{center}
        \hspace{.0\textwidth}\vbox{\hrule\hbox{\vrule\kern6pt
\parbox{.9\textwidth}{\kern6pt#1\vskip6pt}\kern6pt\vrule}\hrule}
        \end{center} \vskip-5mm}
\def\lboxit#1{\vbox{\hrule\hbox{\vrule\kern6pt
      \vbox{\kern6pt#1\vskip6pt}\kern6pt\vrule}\hrule}}

\def\thickboxit#1{\vbox{{\hrule height 1mm}\hbox{{\vrule width 1mm}\kern6pt
          \vbox{\kern6pt#1\kern6pt}\kern6pt{\vrule width 1mm}}
               {\hrule height 1mm}}}

\def\calA{\mathcal{A}}

\def\calC{\mathcal{C}}

\def\calH{\mathcal{H}}
\def\calI{\mathcal{I}}

\def\calN{\mathcal{N}}

\def\calP{\mathcal{P}}

\def\calS{\mathcal{S}}

\def\calX{\mathcal{X}}
\def\calY{\mathcal{Y}}

\def\fat#1{\hbox{\rlap{$#1$}\kern0.25pt\rlap{$#1$}\kern0.25pt$#1$}}

\newcolumntype{R}{@{\extracolsep{0.5cm}}r@{\extracolsep{0pt}}}%
\newcolumntype{E}{@{\extracolsep{0.25cm}}c@{\extracolsep{0pt}}}%

\makeatletter
\newcommand{\distas}[1]{\mathbin{\overset{#1}{\kern\z@\sim}}}%
\makeatother

\newcommand{\blind}{1}

\begin{document}

\if1\blind
{
	\title{\bf Data adaptive covariate balancing for causal effect estimation for high dimensional data}
	\author{Simion De, 
		Jared D. Huling\thanks{corresponding author: huling@umn.edu}\
		\\
		Division of Biostatistics and Health Data Science, University of Minnesota \\ [8pt]
	}
} \fi

\if0\blind
{
	\bigskip
	\bigskip
	\bigskip
	\begin{center}
		{\Large \bf placeholder title}
	\end{center}
	\medskip
} \fi

\date{}

\maketitle

\begin{abstract}
A key challenge in estimating causal effects from observational data is handling confounding and is commonly achieved through weighting methods that balance distribution of covariates between treatment and control groups. Weighting approaches can be classified by whether weights are estimated using parametric or nonparametric methods, and by whether the model relies on  modeling and inverting the propensity score or directly estimates weights to achieve distributional balance by minimizing a measure of dissimilarity between groups. Parametric methods, both for propensity score modeling and direct balancing, are prone to model misspecification. In addition, balancing approaches often suffer from the curse of dimensionality, as they assign equal importance to all covariates, thus potentially de-emphasizing true confounders. Several methods, such as the outcome adaptive lasso, attempt to mitigate this issue through variable selection, but are parametric and focus on propensity score estimation rather than direct balancing. In this paper, we propose a nonparametric direct balancing approach that uses random forests to adaptively emphasize confounders. Our method jointly models treatment and outcome using random forests, allowing the data to identify covariates that influence both processes. We construct a similarity measure, defined by the proportion of trees in which two observations fall into the same leaf node, yielding a distance between treatment and control distributions that is sensitive to relevant covariates and captures the structure of confounding. Under suitable assumptions, we show that the resulting weights converge to normalized inverse propensity scores in the $L_2$ norm and provide consistent treatment effect estimates. We demonstrate the effectiveness of our approach through extensive simulations and an application to a real dataset.

\end{abstract}

\def\spacingset#1{\renewcommand{\baselinestretch}%
	{#1}\small\normalsize} \spacingset{1.5}

\newpage
\spacingset{1.75} 
\setlength{\abovedisplayskip}{7pt}%
\setlength{\belowdisplayskip}{7pt}%
\setlength{\abovedisplayshortskip}{5pt}%
\setlength{\belowdisplayshortskip}{5pt}%



\label{sec:intro}
\section{Introduction}

Estimating causal effects in observational studies is challenging due to confounding. While randomized controlled trials (RCTs) are the gold standard for establishing causality through randomization, they are not always feasible or ethical. As a result, researchers must rely on observational data to investigate critical research questions. However, confounding is common in such studies as individuals self-select into treatment groups, introducing bias and making direct comparisons of outcomes between groups unreliable.  Thus, obtaining unconfounded comparisons between treatment groups is critical when estimating aggregate causal effects such as the average treatment effect (ATE) \citep{kang2007demystifying}. 

Weighting methods are commonly used for confounding control and can generally be classified two different ways. The first classification is whether the weights are derived from a model for the propensity score \citep{rosenbaum1983PropensityScore, robins1995semiparametric, hahn1998role, robins2000marginal} or whether the weights are constructed directly to achieve covariate balance \citep{Chattopadhyay2020BalancingvsWeighting}. The propensity score is the probability that an individual is assigned to treatment given covariates. The propensity score plays an important role in balancing key functions of covariates \citep{imai2014covariate} and this property is used to derive weights that consistently estimate the treatment effect. The propensity score is first estimated by a parametric or non-parametric method and weights are constructed as the inverse of the propensity score when estimating the average treatment effect \citep{rosenbaum1983PropensityScore}. Covariate balancing approaches, on the other hand, aim to reduce confounding bias by ensuring that treatment and control groups are as comparable as possible in their covariate values \citep{ben2021balancing}. The weights are constructed such that some measure of covariate balance between treatment and control groups to the full sample is achieved on the weighted data. 

Both propensity score modeling and covariate balancing approaches can be further categorized as either parametric or nonparametric. For propensity score modeling approaches, this is simply based on whether the model for the propensity score is parametric or not. Nonparametric approaches are often favored, as model misspecification for the propensity score can result in substantial bias \citep{kang2007demystifying}. For covariate balancing approaches, the distinction is less straightforward to characterize, however, approaches that seek balance on a finite set of functions of covariates can be seen as parametric. Such covariate balancing approaches often implicitly model the propensity score \citep{zhao2017entropy, chan2016globally, wang2020minimal} with parametric models with a parametric term for each function of covariates balanced. Nonparametric covariate balancing approaches do not just seek balance on a finite set of functions but either encourage balance of either an infinite dimensional class of functions \citep{hazlett2020kernelbalancing, wong2017kernel, kallus2020generalizedmatching} or the joint distribution of covariates \citep{huling2020energy}. By doing so, nonparametric balancing approaches are less prone to the analyst not balancing on an important function of covariates.

In modern observational studies, the available baseline covariates are often high dimensional. However, most approaches described above do not explicitly aim to deal with high dimensionality.
Although it is straightforward to adapt penalization methods to parametric propensity score modeling approaches, such an approach is still subject to model misspecification. More sophisticated variable selection approaches based on parametric propensity score modeling have been developed. For example, the outcome adaptive lasso method proposed by \citet{shortreed2017outcomeadaptivelasso} uses a penalized logistic regression to model the propensity score wherein the penalty is adaptively modulated based on each covariate's marginal relationship with the outcome. By doing so, the outcome adaptive lasso takes more directly into account the causal structure, emphasizing the selection of true confounders, i.e. variables that affect both treatment and outcome. \citet{tang2023ultra} expand on this to allow for ``ultra''-high dimensional covariates by adding a screening step before the outcome adaptive lasso modeling step. However, these approaches still require the propensity score model to be correctly specified.
Although nonparametric approaches are less susceptible to model misspecification, they are often less resilient to dealing with high-dimensional data.
For example, energy balancing and nonparametric balancing approaches based on maximum mean discrepancy distances with a Gaussian kernel may fail to fully balance the joint distribution in high dimensions, as the distances they are based on fail to characterize joint distributions beyond marginal features in high dimensions \citep{zhu2021interpoint, zhu2020distance, chakraborty2021new, yan2023kernel}. The usage of Euclidean distance, both in the energy distance and in the Gaussian kernel, makes the balancing process prone to the curse of dimensionality. 
The metrics considered in the non-parametric covariate balancing approaches treat each dimension equally and thus do not allow for upweighting true confounders. For example, the energy balancing method \citep{huling2020energy} uses the Euclidean norm to ultimately construct the measure of imbalance, thereby the metric is agnostic to the structure of dependence between covariates, treatment, and outcome and is thus purely design-based.

While non-parametric balancing approaches may be limited in high dimensions, a promising direction for improving them in high dimensions is to adaptively identify true confounders to emphasize distributional balance on them. By doing so, the effective dimensionality may be reduced, potentially dramatically depending on the data at hand.
Considering the limitations of existing methods, we propose a novel approach that has three important characteristics: 1) a weighting approach based on non-parametric covariate balancing, 2) adaptive and flexible identification of confounders, and 3) handling of high-dimensionality. To achieve these three aims, our proposed method involves fitting a multivariate random forest model \citep{bai2022multinomialrandomforest}, treating the outcome and treatment assignment as the dependent multivariate ``response'' and with covariates as predictors. By leveraging the ability of random forests to capture nonlinear and complex dependencies, we design a data-adaptive measure of imbalance using a kernel constructed from the multivariate random forest. In particular,  multivariate random forests are constructed with trees with splits determined by minimizing the average loss across both the prediction of the outcome and the treatment, splits tend to occur in covariates and regions of covariate space that jointly impact outcome and treatment and thus identify confounding structures. As such, we define a measure of distributional balance based on a kernel defined as the probability that two points land in the same leaf node in a randomly chosen tree in the random forest \citep{scornet2016randomforestkernel}. Thus, the distance is more sensitive to confounders and regions in covariate space that have a confounding structure. Further, since random forests are known to adapt to sparsity in the sense that their rate of convergence does not depend on the number of ``noise'' variables \citep{biau2012analysis}, our approach can be used to help mitigate high dimensionality. We demonstrate the data-adaptive nature of our kernel and propose to estimate weights that optimize our metric to obtain weights that minimize the distributional imbalance of confounders. 

Our proposed approach has several additional practical advantages. In particular, since random forests can natively handle mixtures of discrete and continuous covariates, our method can naturally handle mixed types of covariates, unlike distributional balancing approaches which rely on Euclidean distances. Further, because the trees underlying a random forest can handle missing data via methods such as surrogate splitting, our distance and thus weights can be constructed in the presence of missing covariate information.

Theoretically, as an asymptotic theory for random forests with data-adaptive splits is nascent \citep{scornet2015consistency, biau2008consistency, genuer2012variance}, we study the behavior of our weights constructed from a simplified model of a random forest, as is often common in the study of random forests \citep{biau2012analysis}. In particular, we show that a kernel constructed based on random splits is universal and further that weights constructed by minimizing a maximum mean discrepancy with a university kernel converge in an L2 sense to the normalized inverse propensity score.
Furthermore, extensive simulations with varying controlling parameters validate the superior performance of our proposed method compared to existing state-of-the-art approaches.

\section{Preliminaries}
\subsection{Assumptions And Notation}
Let $Y\in\calY$ denote an outcome of interest, $A\in\calA=\{0,1\}$ denote a binary indicator of receipt of treatment, and $X\in\calX$ be a $p$-dimensional vector of pre-treatment covariates. We let
$\{ (X_i,Y_i,A_i)\}_{i=1}^n$ denote an observed dataset where $i$ indexes the units. We further let $S_a=\{i: A_i=a \}, |S_a|=n_a$ for $a=0,1$ represent the set of individuals corresponding to the control and treatment groups, respectively, and $n_a$ for $a=0,1$ represent the number of individuals in these groups. We adopt the potential outcomes framework, where $Y(0)$ is the potential outcome under control and $Y(1)$ is the potential outcome under treatment.  

\begin{figure}[!h]
      \centering
      \includegraphics[width=0.95\textwidth]{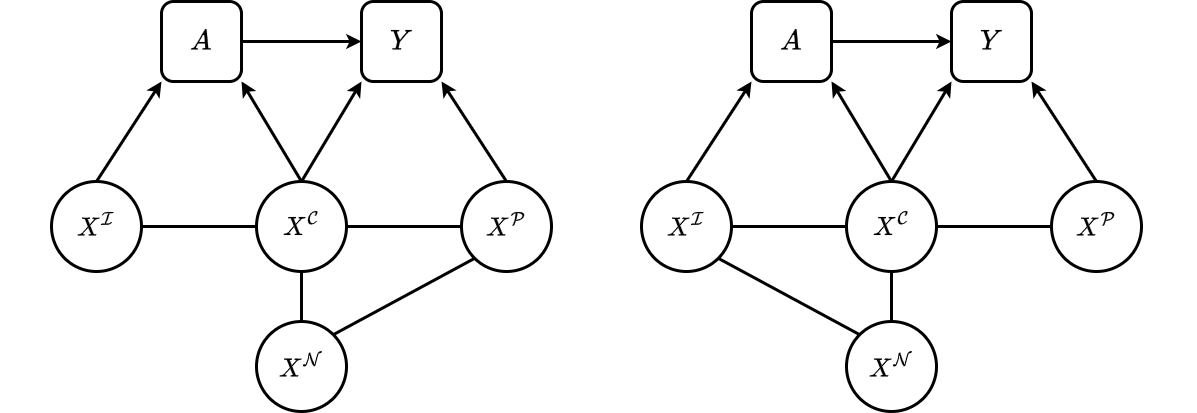}
      \caption{Displayed is a causal directed acyclic graph (DAG) depicting the possible relationships of pre-treatment covariates, treatment assignment, and outcome under our categorization of the different types of pre-treatment covariates. Either DAG is valid for our framework, however we note that there cannot be a complete path between $X^\calI$, $X^\calN$, and $X^\calP$, as this would make $X^\calN$ behave similarly to confounders.}
      \label{fig:causal_dag}
  \end{figure}

 The covariates in $X$ can be grouped into four categories based on their relationship with the treatment and outcome and each other. Loosely-speaking, precision variables affect only the outcome and not the treatment, instrumental variables only affect the treatment assignment and not the outcome, and the confounders affect both the treatment variable and the outcome. Further, null variables are those unrelated to either treatment or outcome. With our categorization of covariates, the covariates can be re-arranged in the following way: $X=(X^\calP, X^\calI, X^\calC, X^\calN)$. Here, $X^\calP$ is the vector of precision variables, $X^\calI$ represents the instrumental variables, $X^\calC$ contains the confounders, and $X^\calN$ is all null variables, i.e. those that have no relationship with any other random variable in the data. The depiction of the (conditional) dependencies that define these four types of covariates are depicted in a directed acyclic graph (DAG) in Figure \ref{fig:causal_dag}. The categorization of covariates can be alternately expressed as statements regarding conditional independencies. Specifically, the precision variables are independent of the treatment assignment conditional on the confounders, i.e. $X^\calP \indep A \;|\; X^\calC$. Instrumental variables are those independent of the outcome conditional on the confounders i.e. $X^\calI\indep Y \;|\; X^\calC$ but that have a relationship with treatment. Here we assume that none of the variables are dependent on the null variables $X^\calN$. Compared with \citep{tang2023ultra}, we consider a relaxed assumption that there can be an association between precision variables, instrumental variables, and confounders. We here assume that null variables are independent of the outcome and the treatment assignments, conditional on confounders and either instrumental variables or precision variables.
 To deal with high-dimensional covariates, we aim to make use of the fact that only confounders must be controlled to mitigate confounding bias and further that precision variables may be controlled for to reduce variance. 
 
In this paper, we use the following standard assumptions, which are common in the causal inference framework.
\begin{enumerate}
    \item \textbf{SUTVA (stable unit treatment value assumption)}: The potential outcomes of each unit are unaffected by the potential outcomes of other units and only one version of the treatment exists. Under SUTVA, the observed outcome is consistent with the potential outcomes in that $Y_i=Y_i(A_i)$. 
    \item \textbf{No Unmeasured Confounding}: We further assume the assignment mechanism is
strongly unconfounded in the sense that $ Y(a) \indep A|X^\calS$ for $a=0,1$ for some $\calS \subseteq \{ 1,2,3,....,p \} $
    \item \textbf{Positivity}: We further assume positivity of the propensity score  $\pi(x^\calS)=P(A=1|X^\calS = x^\calS)$ i.e. $0<\pi(x^\calS)<1$ for all $x^\calS$.
\end{enumerate}

The version of the no unmeasured confounding assumption that we use requires that conditioning on a subset $\calS \subseteq \{ 1,2,3,....,p \}$ of the measured covariates is sufficient to explain away any dependence between potential outcomes and treatment assignments. Positivity requires that any covariate pattern defined by the sufficient subset $\calS$ have a positive probability of being treated or not treated.

 \subsection{A refined error decomposition for weighted estimators of the ATE}
 Here we focus on estimating the average treatment effect (ATE), defined as
$\tau=E[Y(1)]-E[Y(0)]$. We aim to use weighted estimates of the ATE, which take the form  
 $$\hat{\tau}_w=\frac{1}{n_1}\sum_{i=1}^n w_iY_iA_i-\frac{1}{n_0}\sum_{i=1}^n w_iY_i(1-A_i),$$ 
 where $(w_1,w_2,..,w_n)$ is an arbitrary set of sample weights.
 Weighting estimators work by reweighting patients in the treated and control groups to look similar in the weighted data. In doing so, they aim to account for potential bias due to confounding variables. However, in practice practitioners often construct weights that control for all pre-treatment covariates, which may or may not include only true confounders. When there are many pre-treatment covariates, unnecessary adjustment for non-confounders may result in suboptimal performance and large variance.

To address this, we aim to assess the role of weights with respect to covariate balance of different types of pre-treatment covariates. We categorize covariates into precision, confounder, instrumental, and null variables and then aim to mathematically inspect the contribution of each type of covariate to the estimation error of $\hat{\tau}_w$ with arbitrary weights to help motivate weighting estimation approaches that target only critical sources of bias. For this, we introduce the following notation. We let $F_{n, a,w}(x) =\frac{1}{n_a}\sum_{i \in S_a}w_iI(X_i\leq x)$ be the weighted empirical cumulative distribution function (ECDF) of the covariates in the treatment group $A=a$ and $F_n(x)=\frac{1}{n}\sum_{i=1}^nI(X_i\leq x)$ represent the ECDF of the entire sample and let $F(x)$ be the population CDF of the covariates. The weighted ECDF naturally gives rise to the notion of a population weighted distribution. 
We define the implied population weighted conditional distribution function as $F_w(x^\calP,x^\calC|A=a) \equiv E(F_{n,a,w}(x^\calP, x^\calC))$. In other words, we can view $F_{n, a,w}$ as an unbiased estimate of \textit{some} distribution function.
We further let $\mu_a(x)=E(Y^{(a)}|X=x)$ for $a=0,1$ be the conditional mean potential outcome functions for the treatment arms. Based on the classification of variables described in the previous section, we can write $\mu_a(x)=\mu_a(x^\calP,x^\calC)$ since $\mu_a(x)$ is the conditional mean outcome function and the outcome function is only a function of the precision variables and confounders by definition. With this notation, we can  express the error of $\hat{\tau}_w$ as
\begin{equation}\label{eqn:error_decomp_1_2}
     \begin{split}
         \hat{\tau}_w-\tau= {}&  \int_{\calX}\mu_1(x^\calP,x^\calC)d[F_w(x^\calP,x^\calC|A=1)-F(x^\calP,x^\calC)] \\
         &-\int_{\calX}\mu_0(x^\calP,x^\calC)d[F_w(x^\calP,x^\calC|A=0)-F(x^\calP,x^\calC)]\\
        &\quad+\underbrace{\int_{X}\mu_1(x^\calP,x^\calC)d(F_{n,1,w}(x^\calP,x^\calC)-F_w(x^\calP,x^\calC|A=1))}_{\text{mean zero term}}\\
        &\quad-\underbrace{\int_{X}\mu_0(x^\calP,x^\calC)d(F_{n,0,w}(x^\calP,x^\calC)-F_w(x^\calP,x^\calC|A=0))}_{\text{mean zero term}}\\
         &+\underbrace{\frac{1}{n_1}\sum_{i=1}^n w_i\epsilon_iA_i-\frac{1}{n_0}\sum_{i=1}^nw_i\epsilon_i(1-A_i)}_{\text{weighted error term}},
     \end{split}
\end{equation}
which is a refinement of the bias decomposition in  \citep{huling2020energy}. From the above decomposition in equation \eqref{eqn:error_decomp_1_2}, we can see that all the above terms except the first two terms have mean 0 and hence the bias is driven by the first two terms, which are driven by the distributional imbalance of confounders and precision variables, but not instrumental variables. These bias-driving terms are functions of the difference between $F(x^\calP,x^\calC)$, the joint marginal distribution of precision variables and confounders, and $F_w(x^\calP,x^\calC|A=a)$, a conditional weighted distribution of precision variables and confounders. This provides motivation for finding weights that minimize some distance between $F(x^\calP,x^\calC)$ and $F_w(x^\calP,x^\calC|A=a)$, thus reducing magnitude of the bias-driving terms. 
Minimizing the distance between $F(x^\calP,x^\calC)$ and $F_w(x^\calP,x^\calC|A=a)$ is not possible with observed data; however, the distance between their empirical versions can be characterized.  We can thus further decompose the terms driving the bias as
  
The two key terms in \eqref{eqn:error_decomp_1_2} can be expressed as function of the difference between $F(x^\calP,x^\calC)$ and $F_w(x^\calP,x^\calC|A=a)$, which are population quantities. However these corresponding distribution functions can be replaced by empirical versions,  $F_{n,a,w}(x^\calP,x^\calC)$ and $F_n(x^\calP,x^\calC)$, without adding any bias-inducing terms. Thus, the above decomposition thus motivates distributional balancing weights that would minimize the difference between $F_{n,a,w}(x^\calP,x^\calC)$ and $F_n(x^\calP,x^\calC)$. 
  
However, in principle, precision variables \textit{should} only impact variance and not bias, so a closer inspection of their role in the bias term is helpful to better understand their role and the role of confounders in the weighted estimate of the ATE.  To do so, we use a simplified assumption that facilitates presentation.
Under the causal DAG depicted in Figure \ref{fig:causal_dag}, the precision variables given the confounders are conditionally independent of the treatment assignment indicating that $F(X^\calP|X^\calC, A)=F(X^\calP|X^\calC)$, allowing for further simplification of the key bias terms in equation \eqref{eqn:error_decomp_1_2}. We show under some simplifying assumptions that the contribution of precision variables to the bias terms goes away, which we now express as a theorem. We suppose that the weights are constructed as a function of $X$ such that they can be expressed as $w(x)$.

\begin{theorem}\label{Decomposition Theorem_1}
     If the weighting function $w(x)$ used for confounding control is a function of only confounders i.e. $w(x)=w(x^\calC)$ and is a positive real-valued function with $E(w(X^\calC)|A=a)=1$ for $a=0,1$ and if the confounders are continuous with marginal density $f(x^\calC)$ and conditional densities $f(x^\calC|A=a)$ given treatment, then
     \begin{equation}\label{eqn:ConfounderDecomp}
       \begin{split}
&\int_{\calX^\calC}\int_{\calX^\calP}\mu_a(x^\calP,x^\calC)d[F(x^\calP,x^\calC)-F_w(x^\calP,x^\calC|A=a)]\\
&\quad=\int_{\calX^\calC}\int_{\calX^\calP}\mu_a(x^\calP,x^\calC)f(x^\calP|x^\calC)dx^\calP(f(x^\calC)-w(x^\calC)f(x^\calC|A=a))dx^\calC\\
&\quad=\int_{\calX^\calC}\int_{\calX^\calP}\mu_a(x^\calP,x^\calC)dF(x^\calP|x^\calC)d[F(x^\calC)-F_w(x^\calC|A=a)]\\
&\quad=\int_{\calX^\calC}\int_{\calX^\calP}\mu_a(x^\calP,x^\calC)dF(x^\calP|x^\calC)d[F_n(x^\calC)-F_{n,a,w}(x^\calC)]+ R_n+R_{n,w},\\
        \end{split}
     \end{equation}
     
where $w(x^\calC)f(x^\calC|A=a)$ is the conditional weighted density of the confounders and $R_n$ and $R_{n,w}$ are mean zero terms.
\end{theorem}

If we consider the conditional mean of the outcome function given the confounders as $\mu_{a,c}(x^\calC)=\int_{\calX^\calP}\mu_a(x^\calP,x^\calC)dF(x^\calP|x^\calC)$, the mean zero terms can be written as $R_n=\int \mu_{a,c}(x^\calC)d(F(x^\calC)-F_n(x^\calC))$ and $R_{n,w}=\int \mu_{a,c}(x^\calC)d(F_{w}(x^\calC|A=a)-F_{n,a,w}(x^\calC))$. The decomposition in Theorem \ref{Decomposition Theorem_1} shows that under additional assumptions about the weighting function, the distributional imbalance of confounders is the key driver of bias. This motivates our approach to adaptively identify confounders and reduce their distributional imbalance. From Theorem \ref{Decomposition Theorem_1} we can see that the terms driving bias are a function of the difference between $F_n(x^\calC)$ and $F_{n,a,w}(x^\calC)$. Hence, we aim to construct a distributional balancing weight approach that adaptively emphasizes distributional balance on confounding variables but that still allows for balance of precision variables to further reduce variance. 

\subsection{General kernel-based measures for distributional imbalance}
Distributional imbalance drives bias of weighting estimators of the ATE. Recent literature has used distributional distance measures to characterize distributional imbalance after weighting \citep{kallus2020generalizedmatching, huling2020energy, ben2021balancing}. The maximum mean discrepancy (MMD) \citep{borgwardt2006MMD} is a kernel-based measure that can quantify the distributional imbalance using a kernel $K$ in the covariate space $\calX$. Every valid kernel on a space $\calX$ induces a reproducing kernel Hilbert space $\calH$. The $MMD_{\calH}$ is defined as $\sup_{f \in \calH_1}|\int f d(P-Q)|$ \citep{borgwardt2006MMD}. Under the assumption that the function $f \in \calH_1$, where $\calH_1$ is the unit ball in $\calH$ and $P, Q$ are two probability distributions defined on $\calX$ $|\int f d(P-Q)| \leq MMD_{\calH}(P, Q)$, by the definition of $MMD$. We assume that the RKHS induced by the kernel is bounded and the upper bound of the norm for any function in $\calH$ is denoted as the $M$ i.e. $M=\sup_{f \in \calH}\norm{f(\cdot)}$, where $\norm{f(\cdot)}=\sup_{x \in \calX}|f(x)|$. Hence $M$ is related to the kernel and the RKHS induced by the kernel. Thus, under the assumption that the conditional outcome functions $\mu_a(\cdot) \in \calH$ for $a=0,1$, the MMD has the property that 
$$\left\lvert\int_\calX\mu_a(x)d[F_{n,a,w}-F_n](x)\right\rvert \leq M\times MMD_{\calH}(F_{n,a,w},F_n).$$

Thus, MMD serves as a bound on the leading confounding bias terms in \eqref{eqn:error_decomp_1_2} of weighting estimators under a suitable restriction on the space of functions that can characterize the mean outcome functions. As such, minimizing $MMD_{\calH}(F_{n, a,w}, F_n)$ with respect to the weights can also decrease the upper bound on the bias of the weight estimator, provided that $\mu_a(\cdot) \in \calH$. We thus, as in \citet{huling2020energy}  and \citet{kallus2020generalizedmatching}, propose minimizing $MMD_{\calH}^2(F_{n,1,w},F_n) + MMD_{\calH}^2(F_{n,0,w},F_n)$
under the constraint that $\sum_{i \in S_a}w_i=n_a$ to achieve distributional imbalance as well as reduce the bias of the weighting estimator. 

\subsection{Properties of Kernel in High Dimensions}

The choice of kernel in the MMD is critical, as it directly corresponds to how distributional imbalance is characterized. In particular, if one uses a restrictive linear kernel, the MMD will only detect differences in means of covariates. If one uses a kernel corresponding to a particular RKHS, then it characterizes differences in means of functions in the corresponding RKHS \citep{alpay2012RKHS}. Flexible kernels that can detect differences in large function spaces are more robust to complex confounding structures.
Thus, the choice of kernel is of key importance, as it underpins the ability to discern similarities and differences between treatment and control samples, thus facilitating the estimation of treatment effects. However, flexible kernels such as the Gaussian kernel are known to behave poorly in high dimensions \citep{zhu2021interpoint, zhu2020distance, chakraborty2021new, yan2023kernel}, creating a tension between the need to flexibly detect covariate imbalance and the inability to do so well in high dimensions. 
As we have demonstrated in the decomposition in \eqref{eqn:ConfounderDecomp}, balance of the distributions of \textit{confounders} is critical, but imbalance of instrumental variables is unimportant because they do not contribute to bias. On the other hand, the balance of precision variables may improve precision, but will not affect bias. Thus to deal with high-dimensionality, a kernel can be more effective if it de-emphasizes differences in distributions or functions of instrumental and null variables and over-emphasizes imbalance in confounders and potentially on precision variables. 
In this paper, we introduce a kernel formulation that adheres to these principles and, in turn, facilitates a weighting approach that can flexibly control for confounding with a large number of variables.

\section{Methods}
\subsection{A multivariate random forest kernel for high dimensional confounder imbalance}\label{sec:mvar_rf_kernel}
In this section, we propose a kernel that adaptively identifies and emphasizes confounders and regions in covariate space that are confounded in the sense that there is a pronounced joint relationship between covariates, treatment, and outcomes. In doing so, the proposed kernel handles high-dimensionality while balancing confounder distributions. To achieve this, we leverage a multivariate random forest that jointly models the relationship between covariates and treatment assignment and the outcome as a bivariate response. Random forests can be used to construct kernels by defining similarity between two points as the probability that the two points appear in the same leaf node of a randomly-chosen regression tree \citep{scornet2016randomforestkernel, davies2014randompositivedefinite, panda2018learning}. Using a \textit{bivariate} random forest with $(A,Y)$ as the response, trees are split more frequently in covariate space locations that are jointly associated with both treatment and outcome. Thus, the resulting random forest kernel will express similarity of points in terms of how similar they are in terms of their joint relationship with treatment and outcome and thus similarity is measured in terms of the inherent confounding structure in the data. This allows the kernel to more strongly reflect differences in the distributions of confounding variables and in regions in covariate space that have stronger confounding. In the following, we show how such a kernel is constructed, provide an intuition behind why the kernel emphasizes confounders, and demonstrate that it does so with an illustrative toy example.

We use standardized the treatment variable $(\Tilde{A}_i)$ and standardized version of the outcome variable $\Tilde{Y}_i$ jointly as the response variable and the covariates $(X_i)$ as regressors in a multivariate random forest model so that each has mean 0 and variance 1. The use of the standardized version of treatment and outcome instead of the original scales in the multivariate random forest ensures that neither influences the decision trees disproportionately purely due scale differences. As a decision tree divides the covariate space, in our case $\calX$ into several rectangular regions corresponding to the leaf nodes of the decision tree, using the notation of \citet{biau2016random} we define $A_n(x,\theta_j)$ as the region that the value $x\in\calX$ belongs to in the $j$th decision tree in the random forest, where $\theta_j$ is a parameter vector characterizing the $j$th regression tree. Based on the fitted random forest, we define $K_{m,n}(X_1, X_2)=\frac{1}{m}\sum_{j=1}^mI(X_1 \in A_n(X_2,\theta_j))$, keeping the notation consistent with that of \citet{scornet2016randomforestkernel}. With this perspective, the parameter dictating each tree $\theta_j$ is considered random and thus $K_{m,n}(X_1, X_2)$ is an average over the distribution of trees in the forest. The term $K_{m,n}(X_1, X_2)$ is the proportion of times two data points $X_1, X_2$ appear in the same leaf node of trees in a fitted random forest of $m$ trees. Ultimately, $K(X_1, X_2)=P_{\theta}(X_2 \in A_n(X_1,\theta))$ is the stochastic limit of the above term, representing the probability that two points land in the same leaf node throughout the tree distribution. In practice, one cannot fit an infinite number of trees; however, the limiting kernel $K(X_1, X_2)$ can be approximated by growing a large number of trees, e.g. $m=1000$.

To provide intuition behind why using a bivariate random forest can emphasize confounding, we describe the mathematical formulation describing the multivariate random forest and the objective function used to determine splitting points for nodes in the comprising trees. 
Let us consider splitting the criterion used to determine splits for each tree. Define $L$ to be a specific cell or node of the tree and notate the number of data points falling in the cell $L$ as $N_n(L)$. The split criterion for variable $j$, $X^j$ and potential split location $x$ for $X^j$ takes the form  
\begin{equation}
    \begin{split}
        L_n(j,x)= {} &\frac{1}{N_n(L)}\sum\left((\Tilde{Y}_i-\hat{Y}_{L})^2+(\Tilde{A}_i-\hat{A}_L)^2\right)I(X_i \in L)\\
        &-\frac{1}{N_n(L)}\sum\left(\Tilde{Y}_i-\hat{Y}_{<x,L})^2+(\Tilde{A}_i-\hat{A}_{<x,L})^2\right)I(X_i^j<x)\\
        &-\frac{1}{N_n(L)}\sum\left((\Tilde{Y}_i-\hat{Y}_{\geq x,L})^2+(\Tilde{A}_i-\hat{A}_{\geq x,L})^2\right)I(X_i^j\geq x),
    \end{split}
\end{equation}
where $\hat{Y}_{L}$ and $\hat{A}_L$ are the average standardized response and standardized treatment values for observations with $X_i\in L$, respectively, and $\hat{Y}_{<x,L}$ and $\hat{A}_{<x,L}$ are the average standardized response and standardized treatment values for observations with $X_i\in L$ with the additional constraint that $X_i^j<x$, respectively. $\hat{Y}_{\ge x,L}$ and $\hat{A}_{\ge x,L}$ are similarly defined with the additional constraint that $X_i^j\ge x$. 
The form of $L_n(j,x)$ shows that locations $x$ that result in larger improvements in the average mean squared error of the tree's predictions for both treatment and outcome are more likely to result in splits, resulting in nodes that are more homogeneous in terms of their joint predictive relationship between covariates and treatment and outcome. We note that because treatment and outcome are often on different scales, they should be standardized such that they contribute equally to the above loss function. Thus, a random forest of such trees yields a kernel with a value of 1 indicating $X_1, X_2$ are as similar as possible in predicting the treatment and outcome, and a value of 0 indicating maximum dissimilarity in the predictions. As a result, the multivariate random forest kernel aims to capture the structure of the association between the variables via the fitted random forest. The kernel is thus sensitive to similarity and dissimilarity in terms of confounding variables and further tends to be less sensitive to instrumental variables and precision variables and is the least sensitive to null variables. 
Thus, when used in an MMD-based measure of distributional imbalance, a kernel constructed by the bivariate random forest emphasizes distributional imbalance in terms of confounders and in terms of more strongly confounded regions of covariate space.

\subsection{Toy example illustrating confounder-emphasizing properties of proposed kernel}

To visualize the described properties of the kernel and gain intuition for its behavior, we consider the following toy example. We simulated the covariates $X=(X^1,X^2,...,X^p)$ that follow a multivariate normal distribution with an autocorrelation covariance structure i.e. the correlation between $X^i,X^j$ is $(-0.25)^{|i-j|}$. We considered two different dimensions for the covariates with $p=50,200$ to showcase two different scenarios. We simulated the assignment of treatment using the propensity score function $\pi(X)=E(A=1|X)=\text{logit}(.(X^1>.5))$. We  also simulated the outcome variable from a normal distribution with variance 2  and using the following mean outcome function $$E(Y|X,A)=.5I(X^1>.5)+.5I(X^2>.5)$$
 In the simulation setup, 
 \begin{figure}[!h]
      \centering
      \includegraphics[width=0.47\textwidth]{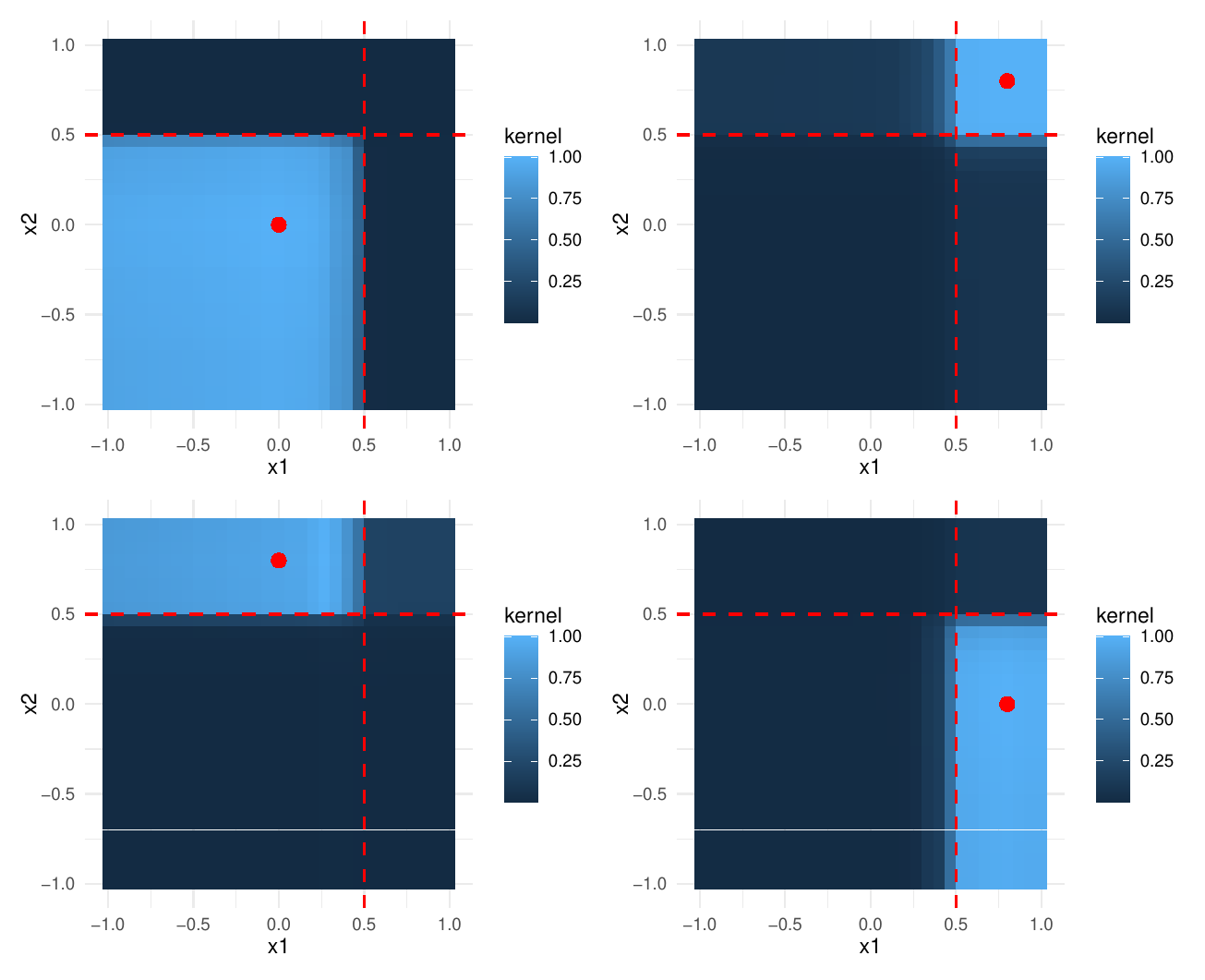}
      \includegraphics[width=0.47\textwidth]{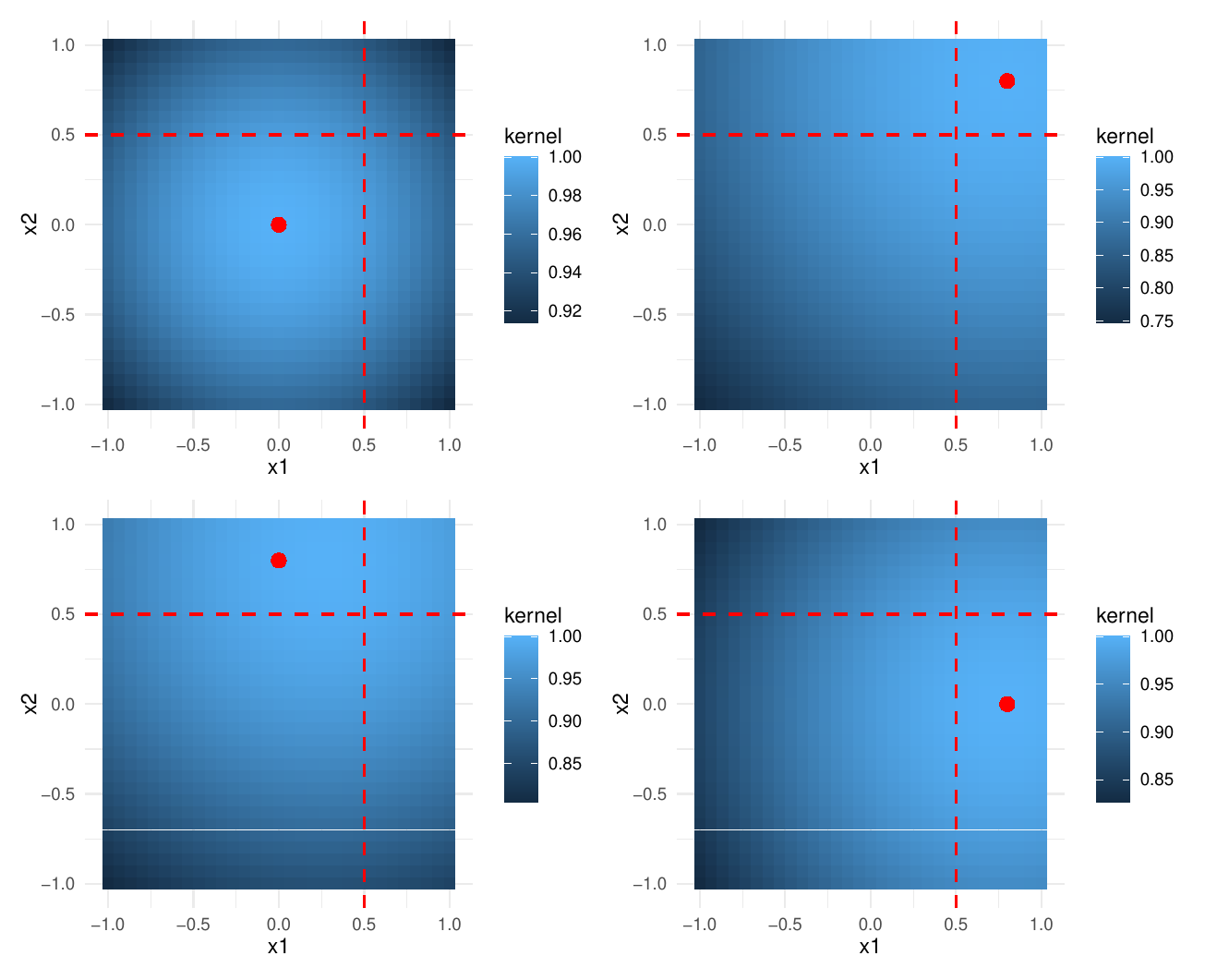}
      \caption{Visualization of our proposed kernel (left 4 plots) and the Gaussian kernel (right 4 plots) plotted on the X-Y plane. The red dots are the centers from which the kernel of every other point has been calculated. The colors are proportional to the values of the kernel, with lighter shades of blue indicating higher similarity and darker indicating lower similarity.}
      \label{fig:kernelVisualization}
  \end{figure}
  the random forest kernel and a Gaussian kernel are displayed with respect to 4 evaluation points in Figure \ref{fig:kernelVisualization}.
 The red dots represent the data points from which the kernel is calculated for every other point. The pattern of the values of our proposed kernel in all 4 left panels in Figure \ref{fig:kernelVisualization} shows prominent 4 regions created by horizontal and vertical lines corresponding to $X^1=.5, X^2=.5$, which are the thresholding values used in the data generation model. The 4 red evaluation points in the 4 regions mentioned above were chosen to emphasize how the kernel varies along $X^1$ (the only true confounder) and $X^2$. Our kernel is more sensitive across $X^1$, which is the only confounder in the data-generating setup. Thus, our kernel emphasizes similarity of points when their inherent confounding structures are similar in the sense that their joint relationship between covariates, outcome, and treatment are similar. Further, the kernel expresses dissimilarity of points when their joint covariate-outcome-treatment relationships are different even if they are similar in covariate-space. The Gaussian kernel, on the other hand, is stationary and radial, exhibiting no preference for either covariate in any location of covariate space. 

In the next section we describe how our kernel is used to construct optimal balancing weights. However, before presenting the details of how such optimal weights are constructed,
we now demonstrate how the similarity properties of our proposed kernel translates to covariate balance when our kernel is used to construct optimal weights in terms of the MMD between treated group and full sample and control group and full sample. We use the same data generation setup used to construct Figure \ref{fig:kernelVisualization}. We then use both our kernel and the Gaussian kernel to construct optimal MMD balancing weights using the same data setup. Further, to compare how our kernel performs compared with inverse propensity score weights, we also include weights based on an estimate of the propensity score by fitting a univariate random forest model with only the treatment assignment as the dependent variable. This comparison shows how our particular use of random forests has advantages over their use in propensity score modeling. We aim to show how well each approach balances the true outcome function, $\mu_1(x)=.5I(x^1>.5)+.5I(x^2>.5)$, as imbalance of this function in particular drives bias. We report the corresponding treatment average and control average as $\frac{\sum_{i \in S_a}w_i\mu_{a}(X_i)}{\sum_{i \in S_a}w_i}$ where $w_i$ are the weights given by three different methods. The three different methods are our method with a random forest kernel (RF Kernel MMD), the energy balancing method with a Gaussian kernel (Gaussian Kernel MMD), and the nonparametric random forest method of estimating the propensity score (RF IPW). \\

For our simulation setup, we have used 500 samples and 50 variables to generate data. That means we are using 48 null variables in our model to have the understanding about performance of our model in presence of null variables. We are incorporating the null variables in our model  to gain insight about the  performance of our method in achieving the distributional balance in the presence of redundant variables. In Table \ref{tab:combined}, we report the population mean, the weighted control mean and the weighted treatment mean, achieved for mean outcome functions $\mu_1=\mu_0$. By population mean(Pop Mean column), we report $\frac{1}{n}\sum \mu_1(X_i)$. In the mean column of treatment, we report $\frac{\sum_{i \in S_1}w_i\mu_1(X_i)}{\sum_{i \in S_1}w_i}$ and in the mean column of control, we present $\frac{\sum_{i \in S_0}w_i\mu_1(X_i)}{\sum_{i \in S_0}w_i}$, where $w_i$ are the weights predicted by the three different methods reported in the table.
From Table \ref{tab:combined} we observe that mean of the outcome function in the treatment and control group after weighting closer to true population mean implying  better balance.
In Table \ref{tab:combined}, we also present similar results, illustrating the balance of the function $\mu_1 = \mu_0$. The analysis uses the same data generation model as in the prior table; however, it incorporates sample sizes equaling 500 and uses a dimensionality of $p=200$. To investigate the impact of high-dimensional settings on balancing properties, we increased only the number of null variables. This adjustment aims to assess whether the inclusion of additional redundant variables enhances the performance of our method relative to existing approaches. 

 \begin{table}[!h]
\centering
\begin{tabular}{l c c c c}
\toprule
\textbf{Design} & \textbf{Process} & \textbf{Pop Mean} & \textbf{Treatment Mean} & \textbf{Control Mean} \\
\midrule

\multirow{3}{*}{\textbf{p=50}} 
 & Gaussian Kernel MMD      & .315 & .327 & .292 \\
 & RF IPW                   & .315 & .335 & .285 \\
 & Proposed RF Kernel MMD   & .315 & .312 & .308 \\

\midrule

\multirow{3}{*}{\textbf{p=200}} 
 & Gaussian Kernel MMD      & .312 & .365 & .214 \\
 & RF IPW                   & .312 & .389 & .205 \\
 & Proposed RF Kernel MMD   & .312 & .344 & .250 \\

\bottomrule
\end{tabular}
\caption{Weighted averages of the outcome function in the treatment and control groups using Gaussian kernel, inverse propensity weighting, and the proposed RF kernel-MMD method, across two simulation scenarios.}
\label{tab:combined}
\end{table}

As evident in Table \ref{tab:combined}, our proposed method achieves the closest alignment between the mean of the outcome function for the treatment and control groups relative to the mean population. Furthermore, the results demonstrate that our method improves its overperformance with respect to the comparator approaches in terms of balance compared with the first scenario, with $p=50$. This improvement suggests that our approach tends to exhibit superior performance as the number of null variables increases.
The results underscore the ability of our method to assign lower weights to redundant variables. By achieving better balance with respect to the mean function, most notably in finite-sample scenarios, our approach is likely to produce more accurate estimates of the average treatment effect.

\subsection{Distributional balancing weights}

In this section, we show how our kernel can be used to construct a measure of distributional imbalance and further formulate an optimization problem for deriving distributional balancing weights. We use the fact that a kernel $K$
on the covariate space induces a reproducing kernel Hilbert space $\calH$. We propose to use the random forest kernel discussed in Section \ref{sec:mvar_rf_kernel} and the RKHS induced by it to define a distance between two probability distributions, in particular a MMD. Due to the equivalence between MMDs and kernel energy distance \citep{sejdinovic2013}, we note that this distance can also be interpreted as a kernel energy distance.

We use the square of the MMD distance between two probability measures to measure the distributional imbalance between the weighted treatment and control distributions and the true population distribution. The square of the MMD can be expressed as a quadratic function of the weights in the following way
\begin{equation} \label{MMD}
\begin{split}
MMD_{\calH}(F_{n,a,w},F_n)^2 =&-\frac{2}{n_an}\sum_{i=1}^n\sum_{j=1}^n w_iI(A_i=a)K(X_i,X_j)\\
&+\sum_{i=1}^n\sum_{j=1}^n\frac{1}  {n_a^2}w_iw_jI(A_i=A_i=a)K(X_i,X_j)\\
&+\sum_{i=1}^n\sum_{j=1}^n\frac{1}{n^2}K(X_i,X_j).
\end{split}
\end{equation}
The main idea is under the assumption that $\mu_1$ and $\mu_0$ in $\calH$
we can say that 
$$\left|\int_{\calH}(\mu_a(x))d[F_{n,a,w}-F_n](x)\right| \leq M* MMD_{\calH}(F_{n,a,w},F_n).$$
  The square of MMD is a valid meaure of distance between two probabilty measure as  \citep{gretton2012kernel} illustrated that if $\calX$ is a compact space, then $MMD_{\calH}$ gives us a metric of probability distribution on $\calX$. Therefore, minimizing the proposed measure of imbalance would perform two tasks. It would reduce the distributional imbalance and, moreover, it would also reduce the terms which are the primary source of bias of the weighting estimator. So following the idea of \citep{Chen_2023balance} and considering the problem of confounder imbalance, we can ultimately write the  optimization problem as 
\begin{equation}\label{objective-function}
	\argmin_{w} MMD_{\calH}(F_{n,1,w},F_n)^2+MMD_{\calH}(F_{n,0,w},F_n)^2+\frac{\lambda}{2}\norm{w}^2
\end{equation}
subject to $\sum_{i \in S_a}w_i=n_a$ $a=0,1$.

The constraints on the sum of $w_i$ ensure that the weighting estimator is normalized and further that $F_{n,a,w}$ are well-defined distribution functions. The penalty $\frac{\lambda}{2}\norm{w}^2$ controls the variability of the weights. In practice, the kernel should be constructed using sample splitting, with one random subsample being used to construct the random forest and the remaining portion of the sample to be used to evaluate the kernel and construct weights by optimizing \eqref{objective-function}.

\subsection{Universality of random forest-style kernels}\label{kernels}

In this section we explore the universality of random forest-based kernels. The universality of a kernel relates to a kernel's ability to represent arbitrary continuous functions and is thus an important property. In a later section, we show that if the weights defined in (\ref{objective-function}) are constructed with a universal kernel, they result in desirable asymptotic properties when estimating the ATE. Thus, to study our proposed kernel, it suffices to explore whether our kernel is universal.
Studying the asymptotic properties of data-adaptive random forests has long been a theoretical challenge given the complexity of the algorithm and its historical resistance to rigorous analytical treatment \citep{cattaneo2025inference}. As such, much work has been devoted to studying simplified models of random forests, such as uniform random partitioning forests, centered forests, Mondrian forests, among many others \citep{biau2016random, cattaneo2025inference}. We follow this track and study kernels formed using uniform random partition forests and further provide a result showing a sufficient condition for any random forest to guarantee its universality, opening the door for the possibility of universality of data-adaptive random forests.

For a uniform random partitioning forests, for each decision tree of the forest a covariate is chosen randomly from all $p$ covariates at each node of the tree and the split is chosen uniformly at random over the range of the selected covariate. This particular process is performed recursively for $k$ steps, where $k$ is pre-chosen parameter. After the process is performed our covariate space is split into $2^k$ rectangular areas and the predicted outcome value at the data point $x$ for a single tree would simply be the average of all the outcome values in the cell of the point $x$. We refer to this simplified version of the random forest as a random forest with uniform random partitioning. The kernel of the random forest partition uniform is defined as $K^*(X_1, X_2)=P(X_2 \in A^*_n(X_1))$ where we define $A^*_n(X)$ as the region of $X$ in the single decision tree in the random forest of a uniform randomly partitioning random forest. To understand universality, we present three conditions that are sufficient to prove that the kernel $K^*$, produced by the uniform random partitioning random forest is universal. We first present the definition of universal kernel.
\begin{definition}[Universal Kernel]
For all compact sets $Z \subseteq \calX$ define the space of all continuous scalar valued functions from $Z \to R$ as $\calC(Z)$ and define $K(Z)=\Bar{span}(K(y,\cdot) :y \in Z)$. The kernel $K(\cdot,\cdot)$ is called universal if for all $Z \subseteq \calX$ and $Z$ compact, $K(Z)$ is dense in $C(Z)$ in the maximum norm i.e. for all $f \in \calC(Z)$, for all $\epsilon > 0$ there exists a $g \in K(Z)$ such that 
$$|| f-g ||_{Z} <\epsilon,$$ 
where $\norm{f}_{Z}=sup_{z \in Z}|f(z)|$.
\end{definition}

We now claim that the three following conditions can be used to  guarantee universality of a kernel.
\begin{description}
    \item[Condition 1:] $K(x_1,x_2)$ is a continuous function for all $(x_1,x_2) \in \calX^2$.
    \item[Condition 2a:] for all $(x_1,x_2) \in \calX^2$ , there exists a measurable set $\Lambda \subseteq \Theta$ such that 
    $\int P(x_1 \in A_n(x_2,\theta))d\mu(\theta)< \mu(\Lambda)$.
    \item[Condition 2b:] for all $(x_1,x_2) \in \calX^2$, there exists a measurable set $\Lambda \subseteq \Theta$ such that $P(x_1 \in A_n(x_2,\theta) <1-\epsilon$  for all $\theta \in \Lambda $ for some $\epsilon > 0$  and $\mu(\Lambda)>0$.
\end{description}

\begin{theorem}\label{Universality Theorem}
   Any kernel $k(\cdot,\cdot)$ that satisfies Condition 1 and either of Conditions 2a or 2b is a universal kernel.
\end{theorem}

The following result shows that the uniform random partitioning random forest kernel is universal.

\begin{prop}\label{Uniform Random Forest Kernel}
 The kernel $K^*(\cdot,\cdot)$, produced by the random forest kernel using uniform random partitioning, satisfies Conditions 1 and 2a above, and hence is a universal kernel.  
\end{prop}

Although our proposal involves using a data-adaptive random forest kernel, Proposition \ref{Uniform Random Forest Kernel} can be viewed as a stepping stone towards understanding the behavior of the data-adaptive random forest used in practice.  Proposition \ref{true random forest} below extends this result to a broader class of partitioning-based kernels that are constructed with trees that have a requirement regarding the properties of the maximum area of their leaf nodes. If we consider $\phi_{w}$ to be the individual trees that comprise a random forest for $w=1,2,3,...,m$, the tree partition area $N(\phi)$ is defined as the largest area of the partitions across the $m$ trees.
\begin{prop}\label{true random forest}
    If the tree partition area converges to zero, i.e. $N(\phi)\rightarrow 0$, the kernel $K$ produced by any, possibly data-adaptive, random forest satisfies Conditions 1 and 2a and is thus universal.
\end{prop}
This Proposition \ref{true random forest} generalizes the universality result for kernels constructed from any random forest which satisfies the condition of tree partition area converging to 0. This constitutes a broader class of random forests as $n$ and $m$, the number of trees, diverge to $\infty$, the trees grow to larger depth and make finer partitions, i.e. the tree partition area converges to 0. 

\subsection{Asymptotic results for balancing weights with a universal kernel}
In this section we present results that describe the asymptotic behavior of the weights {that solve \eqref{objective-function}} under the assumption that the kernel used in the optimization problem satisfies certain conditions. 
In particular, our result requires that the kernel used is universal and positive definite. As a consequence, in order to understand the asymptotic properties of the weights produced by our random forest kernel, it suffices to show whether or not the kernel is universal and positive definite. 
As discussed at length in \citep{biau2016random}, understanding the mathematical properties of the complex data-adaptive random forests has long been a challenge. It is thus common in the literature to explore the mathematical properties of simplified models of random forests to gain insight into the behavior of data-adaptive random forests. In the similar fashion, we consider the kernel produced by a simpler version of a random forest. We consider a random forest where splitting occurs randomly over a range for the covariates with uniform probably and we refer to that as a uniform random partitioning kernel. We can show that the kernel produced in the same manner by this simplified uniform random partitioning kernel is universal and positive definite. 
This result is a key step towards the asymptotic behavior of the balancing weights produced by our proposed kernel. 
Further, we show that any random forest kernel produced by a random forest that has the property that the tree partition area converges to zero also satisfies universality and positive-definiteness, broadening the scope of our results substantially and allowing for the possibility of universality of data-adaptive random forest kernels.
Theorem \ref{The main Result} provides the asymptotic behavior of  balancing weights a universal kernel and shows that such weights converge to the true inverse propensity score weights in the L-2 sense, similar to the results of \citet{Chen_2023balance}.
\begin{theorem}\label{The main Result}
    Suppose that $(\hat{w_i}:1\leq i \leq n)$ be the set of weights that solve the optimization problem \eqref{objective-function}, 
    and the unit ball in $\calH$ is $P-Donsker$ and $\pi(x)$ function is a continuous function and $K(\cdot,\cdot)$ is an universal kernel the following holds:
    \begin{itemize}
        \item The solution for the limiting optimization problem of \eqref{objective-function} as $n\to \infty$, $\hat{w_i}^*$, takes the form $\hat{w_i}^*=A_i\frac{\pi}{\pi(X_i)}+(1-A_i)\frac{1-\pi}{1-\pi(X_i)}$, which is  the inverse of the true propensity score.
        \item Moreover, $\frac{1}{n}\sum(\hat{w_i}-\hat{w_i}^*)^2 \overset{p}{\to} 0$ as $n \to \infty$.
    \end{itemize}
\end{theorem}
The first part of the result is significant to our method's ability to provide balance in terms of mean. As we have shown for $n \to \infty$ the optimal weights are converging to the normalized inverse propensity score thereby balancing the treatment and control mean as well in the long run. Thereby we are achieving both distributional imbalance for finite sample size cases as well as balance in terms of moments in the long run as well. By the second part of the result, we are confirming that the convergence of the weights happens in L-2 norm and that the chance of the L-2 norm being greater than any positive constant value is 0.

\section{Simulation}

\subsection{Simulation Setup}

In this section, we conduct simulation studies to evaluate the performance of the proposed weighting method in finite sample settings. We generate all covariates $X=(X^1, X^2,\dots,X^p)$ from a multivariate normal distribution with an autocorrelation covariance structure with correlation between $X^i,X^j$ as $\rho^{|i-j|}$ with $\rho=.25$ and variance 1. Next, we generate the treatment assignment $A_i$ variable from a $Binomial(\pi(X))$ distribution, and after this generate the outcome using the mean function $\mu_a(x)$ plus an error term drawn from an independent standard normal distribution. We generate data for a variety of sample sizes $n$, and the dimension of covariates $p$ for each different simulation model.

\subsection{Comparator Methods}

For comparison, we considered other weighting methods to show the performance of our method in comparison to other existing state-of-the-art methods. 
We selected 4 major other methods to compare with to highlight the various properties of our method and to assess how it compares with state-of-the-art or standard approaches. 
As a baseline, we compare our method with inverse propensity score weighting. In particular, we compare with both a parametric logistic regression model and non-parametric approach to estimate the propensity scores using random forests.  By comparing with a random forest based approach to IPW, we aim to see how using random forests for the purpose of balancing the covariate distribution compares with using random forests to estimate a propensity score.
We also compare our method with the outcome adaptive lasso, which also performs variable selection to control high dimensionality but assumes linear dependence between variables. By comparing with the outcome adaptive lasso, we aim to see how the issue of model misspecification impacts results even when dimension reduction techniques are used. Moreover, we compare with energy balancing using the Gaussian kernel to perform weighting and compare to our method. Though energy balancing is a non-parametric direct balancing approach, it does not perform variable selection thus targets balance of all variables equally. We present results produced by these methods along with our method for different data generation models in a later subsection. Bias and box plots of the errors of estimates given by the 5 methods: outcome adaptive lasso (OutLasso), inverse propensity score weighting by logistic regression (Logistic IPW), inverse propensity score weighting by random forest (RF IPW), energy balancing with Gaussian kernel (Gaussian Kernel MMD), and finally our proposed method (RF Kernel MMD) are provided.

\subsubsection{Model 1: Smooth nonlinear data generation process}
For Model 1, we generate data to assess the performance of our proposed method under smooth nonlinear associations of covariates with the treatment assignment and the mean outcome function. The data generation uses the function $\pi(x)$ and the outcome function $\mu_a(x)=E(Y|A=a,X=x)$, which are defined as below:
 $$\pi(x)=.25(1+\beta(x^1,2,4)), \text{ and}$$
$$E(Y|A,X=x)=2(x^1-1)+.5(2A-1)[1+1/(1+\exp\{-20(x^1-1/3)\})][1+1/(1+\exp\{-20(x^2-1/3)\})].$$

Under this specification, the true ATE is $\tau \approx 1.82$. For this particular method, we use a non-linear continuous function for the $\pi(x)$ and $\mu_a(x)$ and there are only two confounders. This model uses the beta function, defined as $\beta(a,b)=x^a(1-x)^b$. This particular data generation model tests the performance of our proposed method in the presence of complex non-linear continuous functions. To further illustrate the robustness of the method, we introduce additional null variables, covariates that do not contribute to the association structure into the model. This tests the ability of our method to identify confounders and deal with null variables in the presence of nonlinearities.

\subsubsection{Model 2: Simulation model using similar form of function in the propensity score as well the outcome model}
For Model 2, we generate the data to understand the performance of our methods in complex scenarios where the propensity score function and the mean outcome functions uses higher powers of the covariates and discontinuous functions of covariates. For this model, we generate data using propensity score function $\pi(x)$ and the outcome function $\mu_a(x)=E(Y|A=a,X=x)$, defined as follows:
\begin{equation}
\begin{split}
\pi(x)&= 1/(1 + \exp(-(I(x^1 > 0) + I(x^2 < -0.5) - 0.5I(x^3 > 0 , x^4 < 0)\\
             &+2I(x^4 > 0.5, x^5 < -0.5)-2I(x^1 > 0.5, x^2 < 0.5)+0.5x^4 - 0.5(x^5)^2\\
             & -0.5x^6I(x^7>0) + 0.5x^8  + 0.25(x^9)^2 - 0.25(x^{10}) ^ 2))), \text{ and}
\end{split}
\end{equation}
\begin{equation}
\begin{split}
E(Y|A,X=x)&= 5 I(x^1>0) + 5I(x^2 < -0.5) - 5I(x^3 > 0 , x^4 < 0) + 0.5x^4 - (x^5)^2 \\ 
    &+5I(x^4 > 0.5 , x^5 < -0.5) - 5 I(x^1 > 0.5 , x^2 < 0.5) \\ 
    &-5x^6(I(x^7>0 )+ 0.5x^7) + 0.5x^8 + 0.5(x^9)^2 - 0.5(x^{10})^2.
\end{split}
\end{equation}
For this particular model, the mean outcome function $E(Y|A,X=x)$ does not depend on $A$ and thus $\tau=0$. Moreover, $\pi (x)$ and $\mu_a(x)$ uses the same thresholding functions of covariates, higher powers of covariates, and their interactions, amplifying the negative impact on bias of not fully adjusting for these nonlinear terms. This particular setup tests the ability of our proposed method to effectively handle complex confounding structures where the relationships between covariates, treatment assignment, and outcome are driven by discontinuous and multiplicative dependencies.

\subsubsection{Model 3: Simulation model using linear dependencies}
For Model 3, we simulate data to assess the performance of our method when there are merely linear associations of covariates with the treatment assignment and the outcome variable. This setup is ideal for both the parametric IPW approach and the outcome adaptive lasso. We generate data using the following propensity score function $\pi(x)$ and the outcome function $\mu_a(x)=E(Y|A=a,X=x)$, defined below:
$$\pi(x)=1/(1 + \exp(-.25\sum_{i=1}^{10}x^i-.25\sum_{i=21}^{30}x^i)), \text{ and}$$
$$E(Y|A,X=x)=\sum_{i=1}^{20}x^i.$$

Both $\pi(x)$ and the $\mu_a(x)$ functions are linear in terms of the variables. In this setup 30 covariates impact the data-generating mechanism, with 10 being instrumental variables, 10 being confounders and 10 being precision variables. Again $E(Y|A,X=x)$ does not depend on $A$ and thus $\tau=0$. While our method is designed to perform flexible balancing of confounders understanding, methods that directly use the assumption of linearity are likely to perform best. This simulation setup allows us to assess how much of a penalty our approach pays by being flexible.

\subsection{Results}

In this section we show results for all 5 methods for a range of different values of sample size $n$ and dimensionality of covariates $p$. For each setting, we run the simulation experiment 200 times independently and present box plots of the errors for each method for each simulation replication. 

\begin{figure}[!h]
      \centering
      \includegraphics[width=0.85\textwidth]{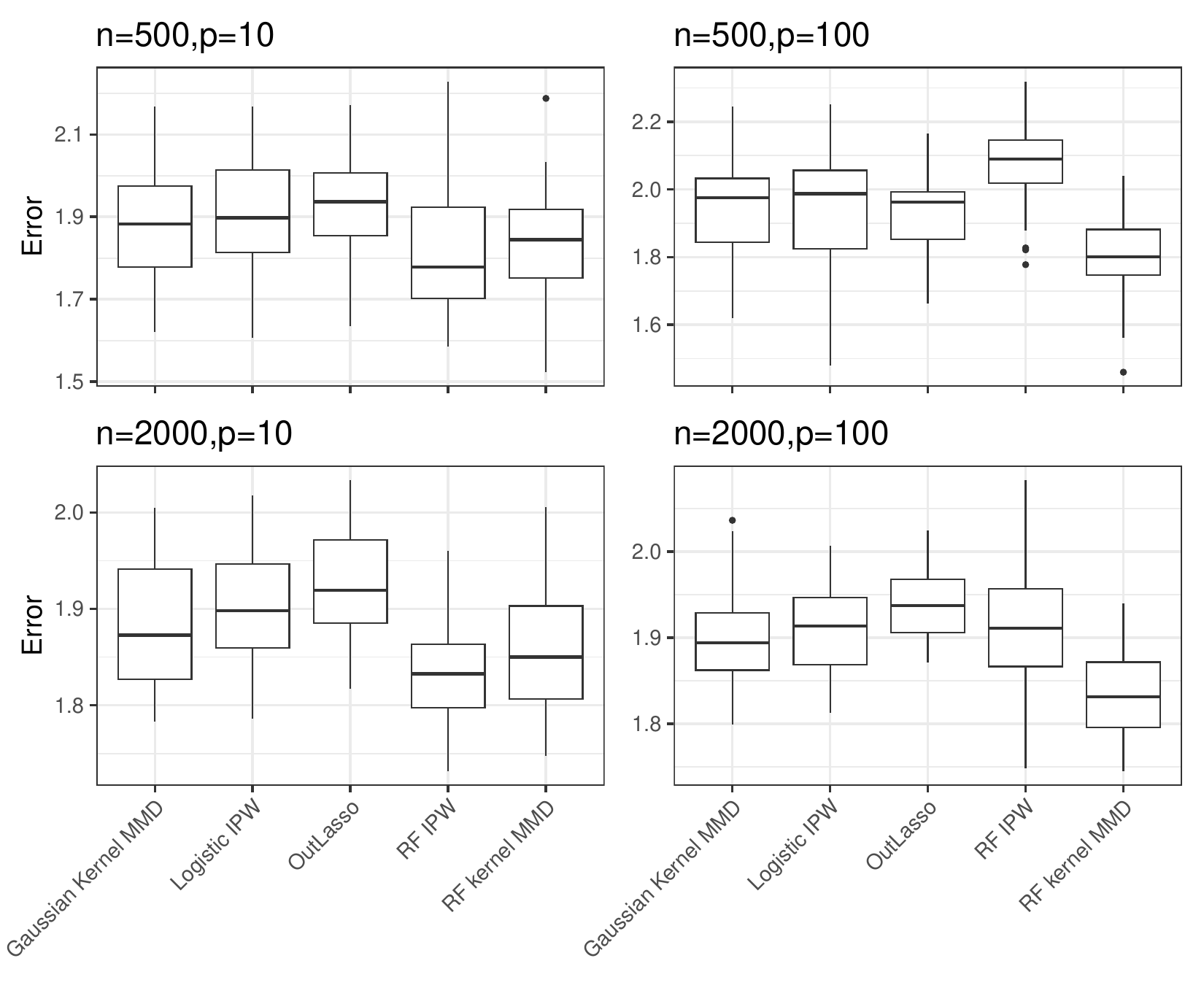}
      \caption{Bias and boxplot of estimates given by 5 methods outcome adaptive lasso (OutLasso), inverse propensity score weighting by logistic regression (logistic IPW), energy balancing with Gaussian kernel (Gaussian Kernel MMD), our method (RF Kernel MMD), inverse propensity score weighting by random forest (RF IPW) for Model 1, which uses smoothly nonlinear data-generating model. }
      \label{fig:bias_results_model_1}
\end{figure}

The results of Model 1, using the smoothly nonlinear data generating model, are displayed in Figure \ref{fig:bias_results_model_1}. From Figure \ref{fig:bias_results_model_1} it can be observed that the bias of the proposed random forest kernel based balancing approach is smaller than for all other approaches and further that the reduction in bias compared with competing approaches is particularly large in the high-dimensional case where $p=100$, when the number of null variables is high. These results suggest our proposed method is able to handle complex, continuous nuisance parameters like the beta function. Further, it is notable that our approach has smaller bias than using a random forest to directly model the propensity score. Thus, the indirect use of the random forest to construct a kernel for the purpose of obtaining balancing weights has a benefit beyond simply flexibly modeling the propensity score.

\begin{figure}[!h]
      \centering
      \includegraphics[width=0.85\textwidth]{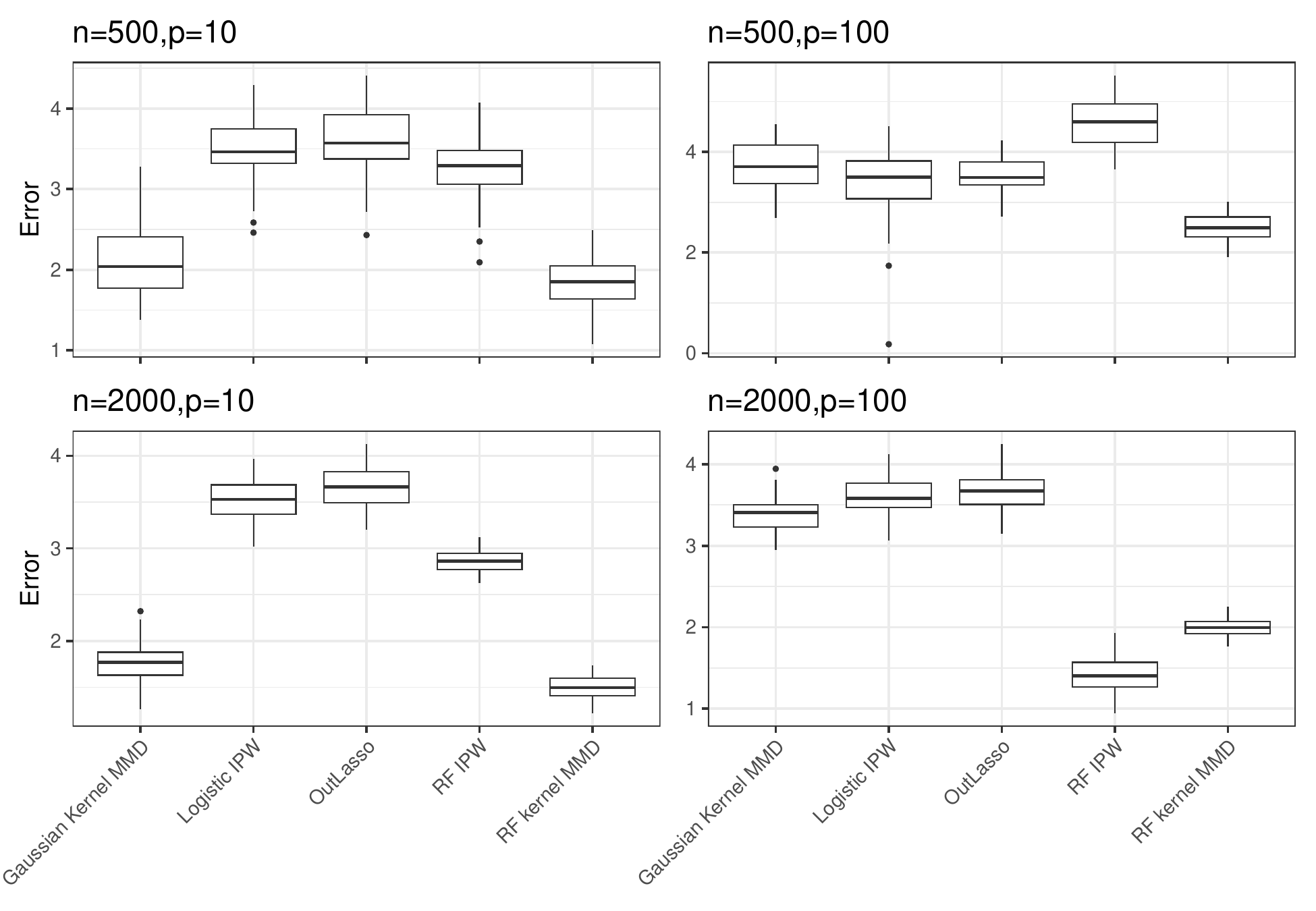}
      \caption{Bias and boxplot of estimates given by 5 methods outcome adaptive Lasso (OutLasso), Inverse propensity score weighting by logistic regression (Logistic IPW), energy balancing with Gaussian kernel (Gaussian Kernel MMD), our method (RF Kernel MMD), inverse propensity score weighting by random forest (RF IPW) for Model 2, which uses same form of function in the outcome model and the propensity score model}
      \label{fig:Highpower}
\end{figure}
The results of Model 2 using higher order terms and discontinuous functions of covariates are shown in Figure \ref{fig:Highpower}. As illustrated in Figure \ref{fig:Highpower}, the bias exhibited by the proposed random forest balancing approach is consistently lower than that of all alternative methods evaluated. Notably, the reduction in bias relative to competing approaches is larger in the high-dimensional scenario, where
$p=100$, particularly when the proportion of null variables is large. These findings indicate that the proposed method demonstrates a robust capacity to accommodate higher-order effects of diverse covariates, demonstrating its effectiveness in complex, high-dimensional settings. 

\begin{figure}[!h]
      \centering
      \includegraphics[width=0.75\textwidth]{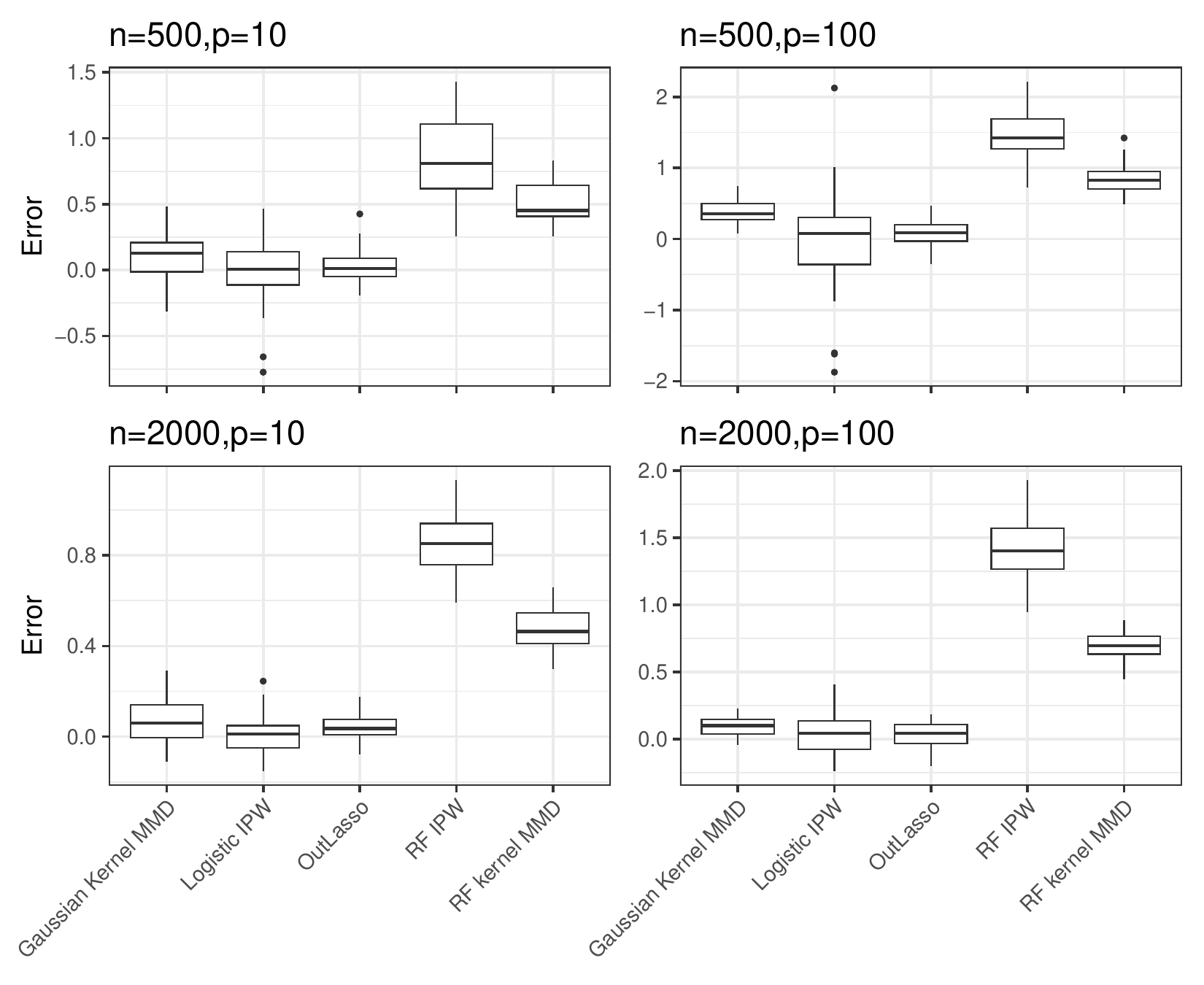}
      \caption{Bias and boxplot of estimates given by 5 methods outcome adaptive lasso (OutLasso), inverse propensity score weighting by logistic regression (Logistic IPW), energy balancing with Gaussian kernel (Gaussian Kernel MMD), our method (RF kernel MMD), inverse propensity score weighting by random forest (RF IPW) for Model 3, which uses linear dependecies}
      \label{fig:linear}
\end{figure}

The results for Model 3 that uses linear form dependencies between variables is presented in Figure \ref{fig:linear}. Our proposed method exhibits slightly larger but comparable bias compared with that of the comparator methods such as the outcome adaptive lasso and the linear propensity score method that are specifically designed to control for linear functions of covariates. This comparability shows the adaptability of our approach in effectively capturing linear relationships, yielding results that remain competitive even under the optimal condition of linearity for competing methods, thereby affirming its versatility across varying association structures.

Collectively, the results suggest that our proposed method demonstrates strong performance in scenarios characterized by high dimensionality and complex data generation processes, thus confirming its adaptability to diverse datasets. Even under ideal conditions (e.g. linearity of relationships) for parametric methods, our method consistently delivers robust performance, yielding reliable and accurate estimates.

\section{Analysis of Right Heart Catheterization Data}

As an illustrative example, we re-analyze observational data from the right heart catheterization (RHC) study of \citet{connors1996effectiveness}. In their original work, \citet{connors1996effectiveness} used a propensity score matching approach to study the effectiveness of RHC in an observational setting, using study data to understand prognoses and preferences for outcomes and risks of treatments. Data were collected on hospitalized adult patients in 5 medical centers in the US. Based on information from a panel of experts, a rich set of variables
related to the decision to perform RHC were collected. The RHC dataset is a collection of information obtained through procedures where a catheter is inserted into the right side of the heart to measure various cardiovascular parameters.

\begin{figure}[!h]
      \centering
      \includegraphics[width=0.7\textwidth]{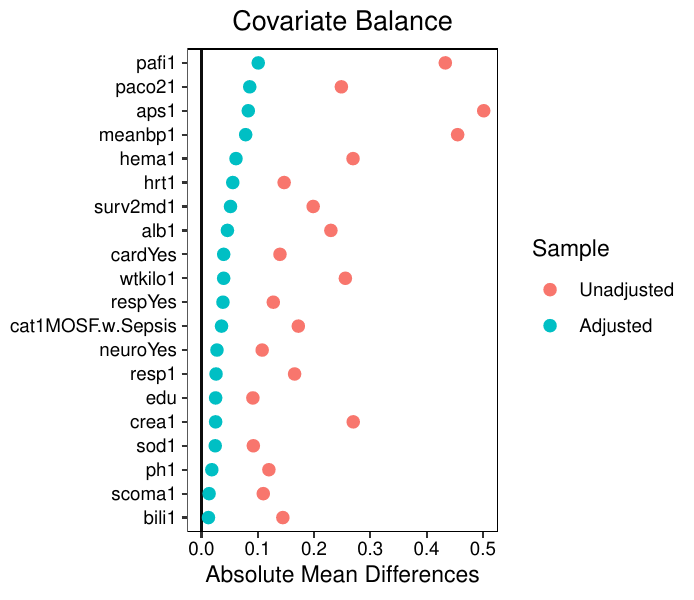}
      \caption{Plot showing standardized mean difference after and before weighting for 20 covariates that shows highest decrease in the standardized mean difference after weighting}
      \label{fig:Balance}
\end{figure}

In our analysis, we use receipt of RHC was used as the treatment and an indicator of mortality within 30 days post-treatment was used as the outcome. For this analysis, we applied our proposed method in addition to the comparator methods explored in our simulation study.  
In order to quantify the degree of balance after weighting for our method, we observed the reduction in standardized mean difference (SMD) before and after weighting. To assess the improvement of the balance of a particular covariate, the reduction in SMD after weighting from SMD before weighting has been used in our analysis. After estimating the weights using our method, we constructed a love plot, shown in Figure \ref{fig:Balance} to compare covariate balance before and after applying the weights estimated by our method. 
As depicted in Figure \ref{fig:Balance}, weighting using our approach resulted in improved balance for several covariates, with this figure illustrating the 20 variables exhibiting the greatest reduction in SMD. Given that our methodology specifically aims to balance confounders, we hypothesized that variables demonstrating the most substantial balance improvement post-weighting are likely confounders, exhibiting significant associations with both treatment assignment and outcome variables.

We further explore whether our approach preferentially improves balance for variables that have a strong relationship with both treatment and outcome, i.e., confounders, compared with variables that have less strong relationships with both or one of the treatment or outcome. A substantial decrease in SMD would indicate that a covariate has a larger improvement in balance after weighting. To assess the strength of relationship with each variable with treatment and outcome, we conducted univariate analysis to examine the associations between these covariates and both the treatment and outcome variables. A strong association with both the treatment and outcome would imply that these covariates are potentially  confounders. 

To evaluate this hypothesis, we performed univariate analyses on all variables, testing their association with treatment and with outcome. For continuous variables, we performed a simple t-test for association between covariates and treatment and between covariates and outcome. For discrete variables we considered a chi-square test the measure these associations. To further explore these relationships, we show a scatter plot, visualizing the test statistics of univariate associations for all variables against their respective SMD reductions. 
\begin{figure}[!h]
      \centering
      \includegraphics[width=0.75\textwidth]{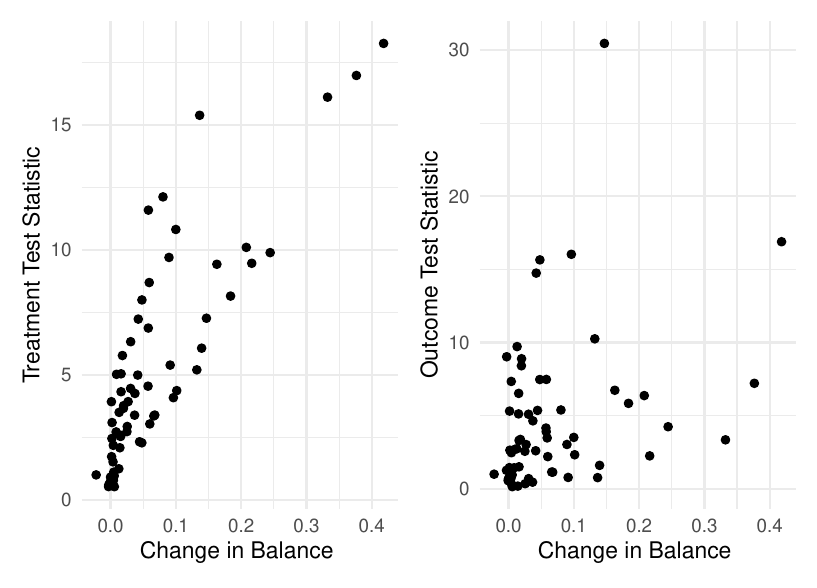}
      \caption{Scatterplot of test statistic of association test of variables with treatment assignment and outcome variable with the change in the standardized mean difference}
      \label{fig:TreatmentBalance}
\end{figure}

In Figure \ref{fig:TreatmentBalance}, we present a graph that illustrates the test statistics for the univariate association of binary categorical treatment and outcome variables with covariates, alongside the decrease in SMD. To assess these associations, we used t-tests for continuous covariates and chi-square tests for categorical covariates, transforming the results into absolute t-statistics and square roots of chi-square statistics for standardized comparison. The figure reveals a trend where increasing balance corresponds to a rough increase in test statistics, indicating that covariates with the greatest SMD reduction show stronger associations with both treatment and outcome, suggesting potential confounding effects, while those with minimal SMD reduction exhibit weaker associations. 
The $p$-values derived from the association tests between the 20 variables shown in Figure \ref{fig:Balance} and both the assignment of treatment and the outcome variables are reported in the Supplementary Materials. The results indicate that the $p$-values for the associations with both the outcome and treatment assignment are consistently low, suggesting that the variables with the greatest balance improvement are indeed potential confounders. This shows the efficacy of our method in adaptively identifying and prioritizing confounders for weighting. Additionally, we investigated variables exhibiting the least reduction in SMD, anticipating weaker associations with either the outcome or treatment assignment. A separate univariate analysis for the 20 variables with the smallest SMD changes with corresponding $p$-values is displayed in the Supplementary Materials. Most $p$-values are higher, typically exceeding 0.05, indicating statistically insignificant associations. This suggests that variables with minimal balance improvement have smaller associations with the outcome and treatment variables, reinforcing the ability of our method to adaptively identify confounders while performing dimension reduction by down-weighting variables with low confounding potential.

    \begin{table}
    \centering
    \begin{tabular}{c c c}
    \toprule
        Method & Estimate & SE \\
        \midrule
        Gaussian Kernel MMD & .0498 & .0124 \\
        RF Kernel MMD & .0484 &.0129\\
        Logistic IPW &.0528 & .0167 \\
        OutLasso & .0548 &.0168 \\
        RF IPW &.0564 & .0121 \\
        \bottomrule
    \end{tabular}
\caption{Table showing mean and standard error of treatment effect estimates for 200 bootstrap samples as calculated by the methods Outcome Adaptive Lasso (OutLasso), Inverse propensity score weighting by logistic regression (Logistic IPW), energy balancing with Gaussian Kernel (Gaussian Kernel MMD), our method (RF Kernel MMD), inverse propensity score weighting by random forest (RF IPW) }
\label{tab:BalanceRealData}
\end{table}

We applied all five methods to the complete RHC dataset to estimate the ATE of RHC on mortality. The application of these methods yielded estimated treatment effect values of 0.0498 (Gaussian Kernel MMD), 0.0484 (Random Forest Kernel MMD), 0.0528 (Logistic IPW), and 0.0548 (Outlasso), 0.0564(RF IPW).  For inference, we used the nonparametric bootstrap for each of five methods using 200 bootstrap resamples to estimate their standard errors. The results from the analysis of the bootstrap samples are summarized in the Table \ref{tab:BalanceRealData}. Notably, our proposed method yields the smallest estimated effect along with the MMD-balancing weights with Gaussian kernel. Thus, both flexible nonparametric balancing methods tend to yield a smaller effect of RHC on mortality than less flexible and/or propensity-score based weighting methods. 

\section{Discussion}

In this paper, we introduced a new method to control for confounding via weighting in observational settings with high-dimensional covariates. Existing methods are either flexible in adjusting for confounding but suffer from the curse of dimensionality, or deal with high-dimensionality directly but can only control for confounding using parametric models that are prone to bias due to model misspecification.
The method introduced in this paper addresses three primary issues: non-parametric covariate balancing, adaptive identification of confounders, and handling high-dimensionality. The proposed method utilizes a multivariate random forest to capture non-parametric dependencies and is used to construct a data-adaptive measure of imbalance using a kernel. We first fit a multivariate random forest model to model the associations between covariates and both the treatment assignment and outcome jointly. Based on this random forest, we then design a kernel that measures similarity of two data points in covariate space based on their similarities in covariate influence on the outcome and treatment jointly, thus emphasizing similarly in terms of confounding relationships. This kernel is in turn used to form a measure of distributional imbalance between two probability distributions. By optimizing this metric, our method achieves flexible covariate balance of confounders. Theoretical analyses support the convergence of weights to the inverse of the propensity score under certain requirements on the random forest. We demonstrated the empirical performance of our proposed method in a suite of challenging simulation studies that emphasizes its good performance especially in high-dimensional settings. Upon submission of our work, we noticed a manuscript \citep{shen2025forest} concurrently posted on arxiv touching on similar issues. The work of \citet{shen2025forest} differs to ours in several important ways. Their approach constructs kernels for balancing using random forests but only consider those constructed on models for the outcome and thus do not share our focus on identifying true confounding structures to balance. Second, this work does not provide asymptotic theory.




\bigskip
\begin{center}
{\large\bf SUPPLEMENTARY MATERIAL}
\end{center}

\begin{description}

\item[Supplementary Material:] The Supplementary Material contains

\end{description}

\def\spacingset#1{\renewcommand{\baselinestretch}%
{#1}\small\normalsize} \spacingset{0.5}

\bibliographystyle{plainnat}

\bibliography{Bibliography}

@article{shen2025forest,
  title={Forest Kernel Balancing Weights: Outcome-Guided Features for Causal Inference},
  author={Shen, Andy A and Ben-Michael, Eli and Feller, Avi and Keele, Luke and Murray, Jared},
  journal={arXiv preprint arXiv:2512.11751},
  year={2025}
}

@article{zhao2017entropy,
  title={Entropy balancing is doubly robust},
  author={Zhao, Qingyuan and Percival, Daniel},
  journal={Journal of Causal Inference},
  volume={5},
  number={1},
  pages={20160010},
  year={2017},
  publisher={De Gruyter}
}

@article{wang2020minimal,
  title={Minimal dispersion approximately balancing weights: asymptotic properties and practical considerations},
  author={Wang, Yixin and Zubizarreta, Jose R},
  journal={Biometrika},
  volume={107},
  number={1},
  pages={93--105},
  year={2020},
  publisher={Oxford University Press}
}

@article{ben2021balancing,
  title={The balancing act in causal inference},
  author={Ben-Michael, Eli and Feller, Avi and Hirshberg, David A and Zubizarreta, Jos{\'e} R},
  journal={arXiv preprint arXiv:2110.14831},
  year={2021}
}

@article{Chen_2023balance,
   title={Robust sample weighting to facilitate individualized treatment rule learning for a target population},
   volume={111},
   url={http://dx.doi.org/10.1093/biomet/asad038},
   DOI={10.1093/biomet/asad038},
   number={1},
   journal={Biometrika},
   publisher={Oxford University Press (OUP)},
   author={Chen, Rui and Huling, Jared D and Chen, Guanhua and Yu, Menggang},
   year={2024},
   month=jun, 
pages={309–329} }

@article{cattaneo2025inference,
  title={Inference with mondrian random forests},
  author={Cattaneo, Matias D and Klusowski, Jason M and Underwood, William G},
  journal={arXiv preprint arXiv:2310.09702},
  year={2025}
}

@article{kallus2020generalizedmatching,
  title={Generalized optimal matching methods for causal inference},
  author={Kallus, Nathan},
  journal={The Journal of Machine Learning Research},
  volume={21},
  number={1},
  pages={2300--2353},
  year={2020},
  publisher={JMLRORG}
}

@article{hazlett2020kernelbalancing,
  title={Kernel balancing},
  author={Hazlett, Chad},
  journal={Statistica Sinica},
  volume={30},
  number={3},
  pages={1155--1189},
  year={2020},
  publisher={JSTOR}
}

@article{scornet2016randomforestkernel,
  title={Random forests and kernel methods},
  author={Scornet, Erwan},
  journal={IEEE Transactions on Information Theory},
  volume={62},
  number={3},
  pages={1485--1500},
  year={2016},
  publisher={IEEE}
}

@article{genuer2012variance,
  title={Variance reduction in purely random forests},
  author={Genuer, Robin},
  journal={Journal of Nonparametric Statistics},
  volume={24},
  number={3},
  pages={543--562},
  year={2012},
  publisher={Taylor \& Francis}
}

@article{biau2008consistency,
  title={Consistency of random forests and other averaging classifiers.},
  author={Biau, G{\'e}rard and Devroye, Luc and Lugosi, G{\"a}bor},
  journal={Journal of Machine Learning Research},
  volume={9},
  number={9},
  year={2008}
}

@article{biau2012analysis,
  title={Analysis of a random forests model},
  author={Biau, G{\'e}rard},
  journal={The Journal of Machine Learning Research},
  volume={13},
  number={1},
  pages={1063--1095},
  year={2012},
  publisher={JMLR. org}
}

@article{scornet2015consistency,
author = {Erwan Scornet and G{\'e}rard Biau and Jean-Philippe Vert},
title = {{Consistency of random forests}},
volume = {43},
journal = {The Annals of Statistics},
number = {4},
publisher = {Institute of Mathematical Statistics},
pages = {1716 -- 1741},
year = {2015},
doi = {10.1214/15-AOS1321},
URL = {https://doi.org/10.1214/15-AOS1321}
}

@article{panda2018learning,
  title={Learning Interpretable Characteristic Kernels via Decision Forests},
  author={Panda, Sambit and Shen, Cencheng and Vogelstein, Joshua T},
  journal={arXiv preprint arXiv:1812.00029},
  year={2018}
}

@article{davies2014randompositivedefinite,
  title={The random forest kernel and other kernels for big data from random partitions},
  author={Davies, Alex and Ghahramani, Zoubin},
  journal={arXiv preprint arXiv:1402.4293},
  year={2014}
}

@article{Chattopadhyay2020BalancingvsWeighting,
author = {Chattopadhyay, Ambarish and Hase, Christopher H. and Zubizarreta, José R.},
title = {Balancing vs modeling approaches to weighting in practice},
journal = {Statistics in Medicine},
volume = {39},
number = {24},
pages = {3227-3254},
doi = {https://doi.org/10.1002/sim.8659},
url = {https://onlinelibrary.wiley.com/doi/abs/10.1002/sim.8659},
eprint = {https://onlinelibrary.wiley.com/doi/pdf/10.1002/sim.8659},
year = {2020}
}

@article{biau2016random,
  title={A random forest guided tour},
  author={Biau, G{\'e}rard and Scornet, Erwan},
  journal={Test},
  volume={25},
  pages={197--227},
  year={2016},
  publisher={Springer}
}

@article{yan2023kernel,
  title={Kernel two-sample tests in high dimensions: interplay between moment discrepancy and dimension-and-sample orders},
  author={Yan, Jian and Zhang, Xianyang},
  journal={Biometrika},
  volume={110},
  number={2},
  pages={411--430},
  year={2023},
  publisher={Oxford University Press}
}

@article{chakraborty2021new,
  title={A new framework for distance and kernel-based metrics in high dimensions},
  author={Chakraborty, Shubhadeep and Zhang, Xianyang},
  journal={Electronic Journal of Statistics},
  volume={15},
  number={2},
  pages={5455--5522},
  year={2021},
  publisher={The Institute of Mathematical Statistics and the Bernoulli Society}
}

@article{zhu2021interpoint,
author = {Changbo Zhu and Xiaofeng Shao},
title = {{Interpoint distance based two sample tests in high dimension}},
volume = {27},
journal = {Bernoulli},
number = {2},
publisher = {Bernoulli Society for Mathematical Statistics and Probability},
pages = {1189 -- 1211},
year = {2021},
doi = {10.3150/20-BEJ1270},
URL = {https://doi.org/10.3150/20-BEJ1270}
}

@article{zhu2020distance,
  title={Distance-based and RKHS-based dependence metrics in high dimension},
  author={Zhu, Changbo and Zhang, Xianyang and Yao, Shun and Shao, Xiaofeng},
  journal={The Annals of Statistics},
  volume={48},
  number={6},
  pages={3366--3394},
  year={2020},
  publisher={JSTOR}
}

@article{huling2020energy,
  title={Energy balancing of covariate distributions},
  author={Huling, Jared D and Mak, Simon},
  journal={Journal of Causal Inference},
  volume={12},
  number={1},
  pages={20220029},
  year={2024},
  publisher={De Gruyter}
}

@article{bai2022multinomialrandomforest,
  title={Multinomial random forest},
  author={Bai, Jiawang and Li, Yiming and Li, Jiawei and Yang, Xue and Jiang, Yong and Xia, Shu-Tao},
  journal={Pattern Recognition},
  volume={122},
  pages={108331},
  year={2022},
  publisher={Elsevier}
}

@book{alpay2012RKHS,
  title={Reproducing kernel spaces and applications},
  author={Alpay, Daniel},
  volume={143},
  year={2012},
  publisher={Birkh{\"a}user}
}

@article{borgwardt2006MMD,
  title={Integrating structured biological data by kernel maximum mean discrepancy},
  author={Borgwardt, Karsten M and Gretton, Arthur and Rasch, Malte J and Kriegel, Hans-Peter and Sch{\"o}lkopf, Bernhard and Smola, Alex J},
  journal={Bioinformatics},
  volume={22},
  number={14},
  pages={e49--e57},
  year={2006},
  publisher={Oxford University Press}
}

@article{tang2023ultra,
  title={Ultra-high dimensional variable selection for doubly robust causal inference},
  author={Tang, Dingke and Kong, Dehan and Pan, Wenliang and Wang, Linbo},
  journal={Biometrics},
  volume={79},
  number={2},
  pages={903--914},
  year={2023},
  publisher={Wiley Online Library}
}

@article{gretton2012kernel,
  title={A kernel two-sample test},
  author={Gretton, Arthur and Borgwardt, Karsten M and Rasch, Malte J and Sch{\"o}lkopf, Bernhard and Smola, Alexander},
  journal={The Journal of Machine Learning Research},
  volume={13},
  number={1},
  pages={723--773},
  year={2012},
  publisher={JMLR. org}
}

@article{shortreed2017outcomeadaptivelasso,
  title={Outcome-adaptive lasso: variable selection for causal inference},
  author={Shortreed, Susan M and Ertefaie, Ashkan},
  journal={Biometrics},
  volume={73},
  number={4},
  pages={1111--1122},
  year={2017},
  publisher={Wiley Online Library}
}

@article{sejdinovic2013,
author = "Sejdinovic, Dino and Sriperumbudur, Bharath and Gretton, Arthur and Fukumizu, Kenji",
doi = "10.1214/13-AOS1140",
journal = "The Annals of Statistics",
month = "10",
number = "5",
pages = "2263--2291",
publisher = "The Institute of Mathematical Statistics",
title = "Equivalence of distance-based and {RKHS}-based statistics in hypothesis testing",
volume = "41",
year = "2013"
}

@article{rosenbaum1983PropensityScore,
  title={The central role of the propensity score in observational studies for causal effects},
  author={Rosenbaum, Paul R and Rubin, Donald B},
  journal={Biometrika},
  volume={70},
  number={1},
  pages={41--55},
  year={1983},
  publisher={Oxford University Press}
}

@article{wong2017kernel,
  title={Kernel-based covariate functional balancing for observational studies},
  author={Wong, Raymond KW and Chan, Kwun Chuen Gary},
  journal={Biometrika},
  volume={105},
  number={1},
  pages={199--213},
  year={2017},
  publisher={Oxford University Press}
}

@article{imai2014covariate,
  title={Covariate balancing propensity score},
  author={Imai, Kosuke and Ratkovic, Marc},
  journal={Journal of the Royal Statistical Society: Series B (Statistical Methodology)},
  volume={76},
  number={1},
  pages={243--263},
  year={2014},
  publisher={Wiley Online Library}
}

@article{chan2016globally,
  title={Globally efficient non-parametric inference of average treatment effects by empirical balancing calibration weighting},
  author={Chan, Kwun Chuen Gary and Yam, Sheung Chi Phillip and Zhang, Zheng},
  journal={Journal of the Royal Statistical Society: Series B (Statistical Methodology)},
  volume={78},
  number={3},
  pages={673--700},
  year={2016},
  publisher={Wiley Online Library}
}

@article{kang2007demystifying,
  title={Demystifying double robustness: A comparison of alternative strategies for estimating a population mean from incomplete data},
  author={Kang, Joseph DY and Schafer, Joseph L},
  journal={Statistical Science},
  volume={22},
  number={4},
  pages={523--539},
  year={2007},
  publisher={Institute of Mathematical Statistics}
}

@article{robins1995semiparametric,
  title={Semiparametric efficiency in multivariate regression models with missing data},
  author={Robins, James M and Rotnitzky, Andrea},
  journal={Journal of the American Statistical Association},
  volume={90},
  number={429},
  pages={122--129},
  year={1995},
  publisher={Taylor \& Francis}
}

@article{hahn1998role,
  title={On the role of the propensity score in efficient semiparametric estimation of average treatment effects},
  author={Hahn, Jinyong},
  journal={Econometrica},
  pages={315--331},
  year={1998},
  volume={66},
  publisher={JSTOR}
}

@article{robins2000marginal,
  title={Marginal structural models and causal inference in epidemiology},
  author={Robins, James M and Hernan, Miguel Angel and Brumback, Babette},
  year={2000},
  journal={Epidemiology},
  pages={550--560},
  volume={11},
  publisher={LWW}
}

@article{connors1996effectiveness,
   author={Connors, Alfred F and Speroff, Theodore and Dawson, Neal V and Thomas, Charles and Harrell, Frank E and Wagner, Douglas and Desbiens, Norman and Goldman, Lee and Wu, Albert W and Califf, Robert M},
  journal={Journal of the American Medical Association},
  title={The effectiveness of right heart catheterization in the initial care of critically Ill patients},
  volume={276},
  number={11},
  pages={889--897},
  year={1996},
  publisher={American Medical Association}
}

\end{document}